\documentclass[final,3p]{elsarticle}

\usepackage{amsmath,amsfonts,bm}

\newcommand{\overbar}[1]{\mkern 1.5mu\overline{\mkern-1.5mu#1\mkern-1.5mu}\mkern 1.5mu}

\def\eqref#1{equation~\ref{#1}}

\def\1{\bm{1}}

\def\vtheta{{\bm{\theta}}}
\def\vepsilon{{\bm{\varepsilon}}}
\def\vsigma{{\bm{\sigma}}}
\def\vphi{{\bm{\phi}}}
\def\vxi{{\bm{\xi}}}
\def\vzeta{{\bm{\zeta}}}

\def\vb{{\bm{b}}}

\def\ve{{\bm{e}}}

\def\vq{{\bm{q}}}

\def\vu{{\bm{u}}}

\def\vx{{\bm{x}}}

\def\ovu{{\overbar{\bm{u}}}}

\def\mC{{\bm{C}}}
\def\mD{{\bm{D}}}

\def\mX{{\bm{X}}}

\def\mPhi{{\bm{\Phi}}}
\def\mPsi{{\bm{\Psi}}}

\def\mSigma{{\bm{\Sigma}}}

\DeclareMathAlphabet{\mathsfit}{\encodingdefault}{\sfdefault}{m}{sl}
\SetMathAlphabet{\mathsfit}{bold}{\encodingdefault}{\sfdefault}{bx}{n}

\def\gG{{\mathcal{G}}}

\def\gL{{\mathcal{L}}}

\def\gN{{\mathcal{N}}}

\def\gS{{\mathcal{S}}}

\def\gV{{\mathcal{V}}}

\def\gZ{{\mathcal{Z}}}

\def\sI{{\mathbb{I}}}

\def\sR{{\mathbb{R}}}

\usepackage{float}
\usepackage[pagewise]{lineno}
\usepackage[dvipsnames]{xcolor}
\usepackage{booktabs}       
\usepackage{framed}
\usepackage{multicol}
\usepackage{amssymb}
\usepackage{todonotes}     
\usepackage[utf8]{inputenc}
\usepackage[ruled, vlined]{algorithm2e}
\usepackage{graphicx}
\usepackage{subcaption}
\usepackage{tabularx}
\usepackage{amsmath}
\usepackage{multirow}
\usepackage{setspace}
\usepackage{microtype}
\usepackage{tabu}
\usepackage{tcolorbox}
\usepackage[export]{adjustbox}
\usepackage[percent]{overpic}
\usepackage{threeparttable}  
\usepackage{tikz}
\usetikzlibrary{arrows.meta}

\DeclareMathAlphabet{\mathcal}{OMS}{cmsy}{m}{n}
\DeclareMathAlphabet\mathbfcal{OMS}{cmsy}{b}{n}

\usepackage{jabbrv}

\usepackage{siunitx}
\usepackage[colorlinks=true]{hyperref}
\usepackage[noabbrev]{cleveref}

\usepackage[normalem]{ulem}
\usepackage{color}

\newcommand{\FEsq}{FE$^{\,2}\,$}
\newcommand{\tr}{\mathrm{tr}\,}

\def\appendixname{\!\!}

\makeatletter
\AddToHook{cmd/appendix/before}{\def\cref@section@alias{appendix}\def\cref@subsection@alias{appendix}}
\makeatother

\journal{Journal of Computational Physics}
\begin{document}
\hypersetup{
    urlcolor=magenta,
    citecolor=blue,
    linkcolor=red
    }
\begin{frontmatter}
\title{EquiNO: A physics-informed neural operator for multiscale simulations}

\author[1]{Hamidreza Eivazi\corref{cor}}
\ead{he76@tu-clausthal.de; hamidreza.eivazi-kourabbaslou@tu-braunschweig.de}
\author[2]{Jendrik-Alexander Tr\"oger}
\author[1]{Stefan Wittek}
\author[2]{Stefan Hartmann}
\author[1]{Andreas Rausch}

\address[1]{Institute for Software and Systems Engineering, Clausthal University of Technology, 38678 Clausthal-Zellerfeld, Germany}
\address[2]{Institute of Applied Mechanics, Clausthal University of Technology, 38678 Clausthal-Zellerfeld, Germany}


\cortext[cor]{Corresponding author}

\begin{abstract}
Multiscale problems are ubiquitous in physics. Numerical simulations of such problems by solving partial differential equations (PDEs) at high resolution are computationally too expensive for many-query scenarios, such as uncertainty quantification, remeshing applications, and topology optimization. This limitation has motivated the development of data-driven surrogate models, where microscale computations are substituted by black-box mappings between macroscale quantities. While these approaches offer significant speedups, they typically struggle to incorporate microscale physical constraints, such as the balance of linear momentum. In this contribution, we propose the Equilibrium Neural Operator (EquiNO), a physics-informed PDE surrogate in which equilibrium is hard-enforced by construction. EquiNO achieves this by projecting the solution onto a set of divergence-free basis functions obtained via proper orthogonal decomposition (POD), thereby ensuring satisfaction of equilibrium without relying on penalty terms or multi-objective loss functions. We compare EquiNO with variational physics-informed neural and operator networks that enforce physical constraints only weakly through the loss function, as well as with purely data-driven operator-learning baselines. Our framework, applicable to multiscale \FEsq computations, introduces a finite element--operator learning (FE-OL) approach that integrates the finite element (FE) method with operator learning (OL). We apply the proposed methodology to quasi-static problems in solid mechanics and demonstrate that FE-OL yields accurate solutions even when trained on restricted datasets. The results show that EquiNO achieves speedup factors exceeding 8000-fold compared to traditional methods and offers a robust and physically consistent alternative to existing data-driven surrogate models.
\end{abstract}


\end{frontmatter}

\section{Introduction}
\label{sec:introduction}

\paragraph{Multiscale modeling} Multiscale systems are defined by exhibiting important features at multiple scales of time and/or space. Such systems arise widely in scientific and engineering domains ranging from heterogeneous solid mechanics \citep{Lin2023,Lee1995}, turbulence in bubble-laden flows \citep{Ma2022}, quantum chemistry \citep{Amaro2018}, granular media and haemodynamics \citep{XIAO2013} to protein conformational dynamics \citep{Alber2019} and astrophysics \citep{Vogelsberger2020}. Multiscale modeling refers to a family of modeling approaches in which multiple models at various scales are simultaneously employed to describe the system, with each typically resolving different levels of spatial resolution. Multiscale models commonly consider physical phenomena on the microscale; however, their examination faces hindrances due to computational effort. The microscopic models often require fine spatial and temporal discretizations to fully resolve the microscale features and capture underlying characteristics, leading to excessively high computational expenses. In recent decades, a range of multiscale methods based on the homogenization theory have been proposed \citep{HASSANI1998,GEERS2010,MATOUS2017} to tackle these difficulties, providing coupled numerical simulations at macroscopic and microscopic scales for the modeling of multiscale systems. 

\paragraph{Multiscale \FEsq\!-method} Multiscale modeling holds particular significance within computational mechanics, especially considering that many widely used engineering materials exhibit heterogeneous microstructures, such as fiber-reinforced polymers or steel alloys. Given the substantial impact of such heterogeneity on material response to mechanical loading, incorporating microstructural considerations into the finite element method (FEM) for structural analyses becomes crucial. Developing constitutive models for materials with heterogeneous microstructures presents challenges in both phenomenological constitutive modeling and subsequent experimental calibration. Hence, the so-called \FEsq$\!$-method has been developed \citep{smitbrekelmansmeijer1998,feyel1999,KouznetsovaBrekelmansBaajiens2001,MieheKoch2002,miehe2003,KouznetsovaGeersBrekelmans2004}. This method enables concurrent numerical simulations of structures at both macro- and microscales using finite elements. In contrast to conventional finite element computations, the \FEsq$\!$-method does not assign a constitutive model to an integration point at the macroscale. Instead, stress and consistent tangent quantities are determined by solving an initial boundary-value problem using finite elements on a specific microstructure, followed by a numerical homogenization technique. This microstructure is typically referred to as a representative volume element (RVE). Alongside the referenced studies, \citet{schroeder2014} offers an extensive overview of the \FEsq$\!$-method for solving coupled boundary value problems across different scales, whereas in \citep{hartmanndileepgrafenhorst2023} a consistent derivation of the algorithmic structure required in such simulations is provided. 

\paragraph{Hybrid AI-based multiscale modeling} More recently, deep-learning (DL) methods have been of interest for multiscale simulations due to their capability to learn complex functions and operators \citep{li2020fourier,lu2021learning,kovachki2023neural} and their flexibility in learning from physical relations \citep{RAISSI2019}. Deep-learning-based surrogates have demonstrated the capacity to simulate partial differential equations (PDEs) up to three orders of magnitude faster than conventional numerical solvers while being more flexible and accurate compared to traditional surrogate models \citep{Karniadakis2021}. The contribution of deep learning in multiscale simulations is mainly in the context of hybrid modeling where neural networks have been employed in conjunction with standard numerical models. Successful applications of deep neural networks for multiscale modeling have been proposed in heterogeneous solid mechanics \citep{Kalina2023,Eivazi2023}, fluid dynamics \citep{Vinuesa2022}, climate \citep{Gentine2018}, and biomedical sciences \citep{Alber2019,Peng2021}.

\paragraph{Physics-augmented surrogate models} There is a relatively new body of research dedicated to augmenting artificial neural networks (ANN)-based constitutive models to fulfill essential physical principles, e.g. thermodynamic consistency, objectivity, material symmetry, and balance of angular momentum (symmetry of the Cauchy stress tensor), by construction \citep{linkaetal2021}. These ANN-based methodologies are recognized in the literature as physics-augmented neural networks (PANNs), as highlighted in \citep{KLEIN2022,Klein2023,LINDEN2023,rosenkranz2024}. This concept traces back to the work by \citep{LING2016jcp,Ling_Kurzawski_Templeton_2016}, where a set of invariant inputs tailored to the problem is constructed. These inputs form the basis for training the ML model, thereby incorporating the invariance directly into the model. Additionally, thermodynamic consistency is achieved by defining the stress as the gradient of a strain-energy density function directly estimated by the ML model. Polyconvexity of this strain-energy density function is ensured through the use of input convex neural networks, as demonstrated in \citep{amos2017}. PANNs have been utilized for multiscale finite strain hyperelasticity problems by \citet{Kalina2023}. Although employing PANNs for multiscale simulations reduces the required number of training samples, the predictions could still violate the governing equations, i.e. the balance of linear momentum. Furthermore, in a multiscale scenario, PANNs do not respect the constitutive relations of the underlying material compositions of the microscale RVE and lead to a substitutive surrogate in which the microscale quantities are not available anymore. This limits the applicability of such methods in inverse microstructure-centered material design and uncertainty quantification.

\paragraph{Physics-informed learning} ML methods have played a revolutionary role in many scientific disciplines, including scientific computing for PDEs \citep{brunton2023machine,Karniadakis2021}. 
Several avenues of research on PDEs have been advanced by machine learning, e.g. learning governing PDEs from data \citep{discovery_pdes}, learning reduced representations \citep{Lee_2020,eivazi_vae}, solving PDEs via Physics-informed neural networks (PINNs) \citep{RAISSI2019,eivazi_pinns_rans,wang2024piratenets}, and finite-dimensional surrogate modeling \citep{Zhu2019surrogate}, among others. PINNs have been utilized for solid mechanics problems related to inversion and surrogate modeling \citep{HAGHIGHAT2021113741,antonetal2025}, solution of PDEs \citep{Samaniego_dem}, and for heterogeneous domains \citep{HENKES2022114790,REZAEI2022115616}. However, most existing approaches are designed for a specific instance of the PDE and can only accommodate a fixed given set of input parameters, or initial and boundary conditions (IBCs), or are mesh-dependent and will need modifications and tuning for different resolutions and discretizations. Recently, a new area of ML research, the so-called operator learning, has significantly advanced the ability to solve parametric PDEs. This approach focuses on learning operators, which are mappings between infinite-dimensional spaces, instead of functions between finite-dimensional spaces \citep{kovachki2023neural}. Operator networks are inherently resolution-independent and can provide solutions for any input coordinate. In contrast to classic PINNs, they can solve new instances of PDEs in a single forward pass without needing additional training.
Notable operator learning methods include DeepONet \citep{lu2019deeponet,lu2021learning} and its extension utilizing proper orthogonal decomposition (POD), referred to as POD-DeepONet \citep{lu2022poddeeponet}, as well as the Fourier neural operator (FNO) \citep{li2020fourier} and PCA-based neural networks (PCANN) \citep{bhattacharya2020model}. Additionally, transformer-based architectures, such as the general neural operator transformer \citep{gnot} and the operator transformer (OFormer) \citep{li2023transformer} as well as graph neural operators \citep{li2020neuraloperatorgraphkernel}, have demonstrated strong potential for capturing complex dependencies in PDE solutions. 

An emerging direction in physics-informed operator learning involves leveraging the variational (weak) form of governing equations to reduce differentiability requirements and improve numerical stability \citep{Kharazmi_vpinn,Goswami_VDeepONet}. The deep energy method (DEM) \citep{Samaniego_dem} directly minimizes the potential energy functional for solid mechanics problems, bypassing strong-form PDE constraints. DEM-based extensions address complex geometries via domain decomposition \citep{Wang_cenn}, and nonlinear contact mechanics \citep{BAI2025}. The variational physics-informed neural operator (VINO) combines the strengths of neural operators with variational formulations to solve PDEs more efficiently. By discretizing the domain into elements, VINO facilitates the evaluation of governing equations where each element's contribution to the variational energy is computed analytically similar to FEM, eliminating reliance on automatic differentiation \citep{Eshaghi_vino}.

\paragraph{Our Contribution}
In this paper, we introduce the Equilibrium Neural Operator (EquiNO), a novel physics-informed neural operator designed to predict microscale fields in multiscale problems. EquiNO provides a \emph{complementary} approach that integrates seamlessly with the multiscale \FEsq framework. Our approach addresses the limitations of traditional surrogate models by adhering to microscale physical constraints, i.e. kinematic and constitutive relations, the balance of linear momentum, and boundary conditions. This methodology enables model training based on physics, without reliance on large datasets. EquiNO approximates RVE solutions by projecting governing equations onto a set of divergence-free POD modes derived from a small dataset, thus forming an efficient reduced-order model for fast inference. EquiNO inherently preserves equilibrium and enforces periodic boundary conditions as hard constraints. In addition to EquiNO, we investigate the use of variational physics-informed operator networks that simulate microscale mechanics using the weak form of the PDE as a loss function, and compare them with state-of-the-art data-driven operator-learning baselines. The proposed methods are evaluated on three benchmark problems and four representative microstructures, including a three-dimensional RVE.

\paragraph{Structure of the paper}
This article is organized as follows: in $\S$\ref{sec:multiscale-simulations}, we provide an overview of the theoretical background relevant to multiscale \FEsq computations. In $\S$\ref{sec:operator-learning}, we review the fundamentals of the physics-informed operator learning, whereas in $\S$\ref{sec:methodology}, we introduce EquiNO. The performance of the methods and their characteristics are discussed in $\S$\ref{sec:results-and-discussion}. Finally, in $\S$\ref{sec:conclusions}, we provide a summary and the conclusions of the study. The source code, data, trained models, and supplementary materials associated with this study can be accessed at our GitHub repository: \url{https://github.com/HamidrezaEiv/EquiNO}.

\section{Multiscale FE$^{2}$ computations}
\label{sec:multiscale-simulations}

This work employs finite element methods for multiscale computations, as detailed in \citep{MieheKoch2002}, \citep{schroeder2014} and \citep{hartmanndileepgrafenhorst2023}, to integrate the effective constitutive behavior of heterogeneous microstructures into macroscale analyses. At the microscale, periodic displacement boundary conditions are applied, aligning with the discussions in \citet{MieheKoch2002}. The process involves concurrent multiscale \FEsq\ computations, where macroscale strains, denoted as $\hat{\ve}^{j}$, impose boundary conditions on the surface of the RVE, representing the microstructure. Subsequently, a boundary value problem (BVP) is solved at the microscale, followed by homogenization to derive the macroscopic stresses $\hat{\vsigma}^{j}$ and the consistent tangent matrix $\hat{\mC}^{j}$. Here, $\left(\cdot\right)^{j}$ indicates quantities associated with macroscale integration point $j$ of an element. Non-linear equation systems are addressed using a Multilevel-Newton algorithm. The study operates in a quasi-static regime, assuming small strains, and uses the principle of virtual displacements. Spatial discretization transforms continuous volumes and surfaces into finite element representations using shape functions, which accommodate both known and prescribed nodal displacements. At the macroscale, equations derive from the discretized balance of linear momentum, while at the microscale, RVEs replace traditional constitutive models, contributing to macroscale properties through homogenization. This establishes a coupling between the macro- and microscale displacements. For a detailed discussion of the \FEsq method, readers are referred to our previous work \citep{Eivazi2023}.

\section{Physics-informed operator learning}
\label{sec:operator-learning}

Our objective is to develop a physics-informed learning technique that acts as a PDE surrogate for microscale physics. In this section, we outline the fundamentals of the operator-learning task and the concept of physics-informed operator learning applied to PDEs. For an in-depth discussion, readers are referred to the literature on the topic \citep{lu2021learning,Wang2021,kovachki2023neural,Goswami2023}.

Consider two separable infinite-dimensional Banach spaces, $\gV$ and $\gS$, over bounded domains. Let $\gN(v, s) = 0$ be a differential operator, which may be linear or nonlinear, and $\gG: \gV \mapsto \gS$ be the solution operator. We consider partial observations, given as input-output data pairs $\{v_i, s_i\}_{i=1}^N$. These pairs satisfy
\begin{equation}
   \gN(v_i, s_i) = 0 \quad \text{and} \quad \gG(v_i)(\vx) = s_i(\vx) \quad \text{ for }i= 1, \ldots, N,
\end{equation}
for any point $\vx$. Here, $v_i \in \gV$ and $s_i \in \gS$. The function $v$ is defined on $\Gamma \subset \mathbb{R}^r$, and the function $s$ is on $\Omega \subset \mathbb{R}^d$. In the context of PDEs, $v$ usually represents the IBCs or parameters, while $s$ denotes the unknown solutions. 
Based on this setup, the goal in a data-driven learning process is to approximate the operator $\gG$ using a dataset $\{v_i, s_i\}_{i=1}^N$.

We focus on DeepONet architecture, as introduced by \citet{lu2019deeponet}. A DeepONet comprises a trunk network and a branch network. The trunk network is responsible for receiving coordinate inputs and producing a set of basis functions. The branch network processes a discretized function $v$ and outputs the coefficients corresponding to the basis functions generated by the trunk network. In a data-driven scenario, the operator $\gG$, which maps the input function $v$ to the output function $s$, is approximated through the linear reconstruction of the output function given by
\begin{equation}
   \gG(v)(\vx)\approx \sum_{k=1}^{p} b_k(v) t_k(\vx) + b_0,
\end{equation}
for any point $\vx$ in $\Omega$. Here, $b_0 \in \mathbb{R}$ represents a bias term; $\{b_1, b_2, \ldots, b_p\}$ are the $p$ outputs from the branch network, and $\{t_1, t_2, \ldots, t_p\}$ are the $p$ basis functions provided by the trunk network. The trunk network inherently learns these basis functions for the output $s$ during the training process.

In the POD-DeepONet \citep{lu2022poddeeponet}, the trunk network is substituted with a set of POD bases while the branch network learns the associated coefficients. The output expression thus becomes
\begin{equation}
   \gG(v)(\vx)\approx \sum_{k=1}^{p} b_k(v) \phi_k(\vx) + \phi_0(\vx),
\end{equation}
where $\{\phi_1, \phi_2, \ldots, \phi_p\}$ denote the POD basis of $s$, and $\phi_0$ is the mean function. Any $\phi(\vx)$ for a point $\vx$ within $\Omega$ can be obtained through interpolation \citep{bhattacharya2020model}. Empirical studies have shown that POD-DeepONets often surpass DeepONets in benchmark operator learning tasks \citep{lu2022poddeeponet}. Training a DeepONet or POD-DeepONet involves minimizing the loss
\begin{equation}
    \label{eq:loss_s}
    \gL(\vtheta) = \dfrac{1}{Nm}\sum_{i=1}^N \sum_{j=1}^{m} \left | \gG_{\vtheta}(v_i)(\vx_j) - s_i(\vx_j)\right |^2,
\end{equation}
where $\gG_{\vtheta}$ represents the deep operator network, and $\vtheta$ encompasses all trainable weights and biases in the model. $N$ is the number of input functions $v$ sampled from $\gV$, and $m$ shows the number of collocation points within the domain $\Omega$.

Although DeepONets have demonstrated impressive potential in learning operators, these models generally require large training datasets consisting of paired input-output observations, which can be costly to acquire. Furthermore, their predictive outcomes may deviate from the fundamental physical principles governing the observed data.~\citet{Wang2021} introduced a novel model category termed physics-informed DeepONets, following the same idea in PINNs. This model class combines data measurements and the principles of physical laws by penalizing the residuals of partial differential equations within the loss function of a neural network, utilizing automatic differentiation. The composite loss function of this model class can be written as
\begin{equation}
    \label{eq:total_loss}
    \gL(\vtheta) = \alpha \gL_s(\vtheta) + \beta \gL_f(\vtheta) + \gamma \gL_c(\vtheta),
\end{equation}
where $\gL_s(\vtheta)$ represents the data measurements loss and is defined as in \cref{eq:loss_s} and 
\begin{subequations}
\begin{equation}
    \label{eq:loss_f}
    \gL_f(\vtheta) = \dfrac{1}{N^f m^f}\sum_{i=1}^{N^f} \sum_{j=1}^{m^f} \left | \gN(v^f_i,\, \gG_{\vtheta}(v^f_i)(\vx_j^f)) \right |^2,
\end{equation}
\begin{equation}
    \label{eq:loss_c}
    \gL_c(\vtheta) = \dfrac{1}{N^f m^c}\sum_{i=1}^{N^f} \sum_{j=1}^{m^c} \left | \gG_{\vtheta}(v^f_i)(\vx_j^c) - \gG(v^f_i)(\vx_j^c) \right |^2,
\end{equation}
\end{subequations}
are the loss components for enforcing the given physical constraints in the form of a system of PDEs and their IBCs, respectively.~$\{v^f_i\}_{i=1}^{N^f}$ denotes a set of $N^f$ input functions sampled from $\gV$, and $\{\vx^f_i\}_{i=1}^{m^f}$ is a set of collocation points randomly sampled from the domain $\Omega$. The superscript $f$ indicates that these input functions and collocation points are utilized to penalize the residuals of the PDEs.~$\{\vx^c_i\}_{i=1}^{m^c}$ denotes a set of collocation points randomly sampled from the domain boundaries $\partial \Omega$. In \cref{eq:total_loss}, $\alpha$, $\beta$, and $\gamma$ are the weighting coefficients for the loss components. For the sake of brevity, the loss terms associated with the initial and boundary conditions are collectively denoted by $\gL_c(\vtheta)$, although they can be treated separately and assigned different weighting coefficients.

\section{Physics-informed learning for continuum micromechanics}
\label{sec:methodology}

In this section, we present our methodology for learning the solution operator of the microscale mechanics in a physics-informed fashion for heterogeneous RVEs. Later, we outline the integration of this acquired operator with the finite element method for conducting multiscale simulations. We formulate our methodology for nonlinear elastostatic problems and propose possible extensions toward path- and rate-dependent problems such as plasticity, viscoelasticity, and viscoplasticity.

\subsection{Governing equations of micromechanics}

The governing equations concerning the microstructure mechanics are discussed in the context of \FEsq~calculations. In contrast to conventional finite element simulations where a constitutive model is assessed at each integration point, here, an initial-boundary-value problem is solved for the evaluation of the microstructure. Subsequently, this is followed by a numerical homogenization, leading to stress and consistent tangent quantities. The PDEs discussed in this work arise in the context of elastostatic problems with nonlinear elastic material behavior. Our focus here is limited to periodic displacement boundary conditions at the microscale and we refer to \citet{MieheKoch2002} for other boundary conditions. Let us consider the two-dimensional domain of interest $\Omega$ to be a symmetric and zero-centered unit cell such that
\begin{equation}
    \Omega=\left\{\vx=\left\{x_1,x_2\right\}^T \in \mathbb{R}^2 \left\lvert\, \frac{-L}{2} \leq x_i \leq \frac{L}{2}, i=1,2 \right.\right\},
\end{equation}
for any point $\vx$ where $L$ indicates the edge length of the domain $\Omega$. We consider periodic displacement boundary conditions on so-called conforming spatial discretizations. These conditions entail coupling the displacements of nodes located across various regions of the boundaries of the RVE. The boundary of the domain is decomposed into opposing boundaries
\begin{equation}
    \partial \Omega = \partial \Omega^+ \cup \partial \Omega^-,
\end{equation}
where two points $\vx^+ \in \partial \Omega^+$ and $\vx^- \in \partial \Omega^-$ are linked by periodicity
\begin{subequations}
    \begin{equation}
        \vu(\vx^+) - \vu(\vx^-) = \left[\mD(\vx^+) - \mD(\vx^-)\right]^T \hat{\ve},
        \label{eq:bc}
    \end{equation}
    \begin{equation}
        \mD(\vx) = \dfrac{1}{2} \begin{bmatrix}
            2\:\! x_1 & 0 \\
            0 & 2\:\! x_2 \\
            x_2 & x_1 
        \end{bmatrix}.
        \label{eq:D}
    \end{equation}
\end{subequations}
Here, $\vu(\vx^+)$ and $\vu(\vx^-)$ are displacement vectors at two linked points $\vx^+$ and $\vx^-$.~$\mD(\vx^+)$ and $\mD(\vx^-)$ are coordinate matrices according to \cref{eq:D} for a two-dimensional domain, and $\hat{\ve}$ is the prescribed macroscale strain tensor in Voigt notation. Following \citep{hauptbuch2}, the governing equations for nonlinear elastostatic problems include the balance of linear momentum 
\begin{equation}
    \label{eq:balanceMom}
    \vsigma(\vx) \nabla = \bm{0}, \quad \vx \in \Omega,
\end{equation}
neglecting body forces and dynamic terms, the kinematic relation 
\begin{equation}
    \label{eq:kinematic}
    \vepsilon(\vx) = \dfrac{1}{2} (\vu(\vx) \otimes \nabla + \nabla \otimes \vu(\vx)), \quad \vx \in \Omega,
\end{equation}
that relates the displacement vector $\vu(\vx)$ to the linearized strain tensor $\vepsilon(\vx)$, and a nonlinear constitutive relation $\bm{h}$
\begin{equation}
    \vsigma(\vx) = \bm{h}(\vepsilon(\vx)),
\end{equation}
describing the dependence of the stress tensor $\vsigma(\vx)$ on the strain tensor $\vepsilon(\vx)$. The balance of linear momentum and kinematics hold similarly for the macroscale problem, as discussed in \citep{Eivazi2023}.

\subsection{Proper orthogonal decomposition}
\label{sec:pod}
In this study, we will develop our complementary learning-based PDE surrogate of the microscale based on the POD-DeepONet architecture \citep{lu2022poddeeponet}. POD is a technique for modal decomposition employed to identify key features from high-dimensional vector fields \citep{Lumley1967,Taira_2017,Taira_2020}. Let us consider a vector field, e.g., the displacement vector field or the stress tensor field in Voigt notation, on a set of collocation points $\vxi$ for a prescribed global strain $\hat{\ve}$. We define $\vq(\vxi, \hat{\ve})$ to be a column-vector representation of the vector field such that the scalar fields of vector components are concatenated into one large column $[\vq_1(\vxi, \hat{\ve}), \dots, \vq_d(\vxi, \hat{\ve})]^T \in \sR^{(m \times d)}$ for $d$ components of the vector field and $m$ collocation points. The column-vector field can be decomposed as 
\begin{equation}
    \vq(\vxi, \hat{\ve}) = \sum_{j=1}^p a_j(\hat{\ve})\vphi_j(\vxi),
\end{equation}
where $\vphi_j(\vxi)$ and $a_j$ represent the $j$th spatial mode and its coefficient, respectively. To achieve this, we first construct snapshots of the vector field as a collection of the column vectors
\begin{equation}
    \vq(\vxi, \hat{\ve}^i) \in \sR^{(m \times d)}, \quad i = 1, 2, \ldots, N,
\end{equation}
where $N$ is the number of snapshots. We arrange the data into a matrix $\mX$ by concatenating $N$ snapshots
\begin{equation}
    \boldsymbol{X} = [\vq(\vxi, \hat{\ve}^1)~\vq(\vxi, \hat{\ve}^2)~\ldots~\vq(\vxi, \hat{\ve}^N)]\in \mathbb{R}^{(m\times d)\times N}.
\end{equation}
The POD modes can be determined by applying singular-value decomposition (SVD) \citep{Sirovich_1987} on the snapshots matrix $\mX$ as
\begin{equation}
    \mX = \mPhi\, \mSigma\, \mPsi^T,
\end{equation}
where $\mPhi \in \sR^{(m \times d) \times N}$ and $\mPsi \in \sR^{N \times N}$ correspond to the left and right singular vectors of the matrix $\mX$, respectively, and $\mSigma\in \sR^{N \times N}$ represents a diagonal matrix comprising the singular values. We define the POD subspaces as
\begin{equation}
    \gZ = \rm{span}\{\phi_1, \phi_2, \dots, \phi_p\},
\end{equation}
for any $p \geq 1$ and leverage them within the proposed operator-learning framework.

\subsection{Equilibrium neural operator (EquiNO)}

In this section, we introduce the equilibrium neural operator (EquiNO), our unsupervised methodology for learning the solution operator of nonlinear elastic RVEs. Our method solves the governing equations for any given prescribed global strain $\hat{\ve}$ by projecting onto a set of POD modes derived from a limited and unpaired dataset. The performance of our method is compared to that of physics-informed models based on variational energy-based formulation and data-driven models, in terms of both accuracy and computational efficiency. It is important to note that physics-informed models do not require solution data and operate in a completely unsupervised manner. We initially focus on solving the nonlinear elastic problem for a specific boundary condition induced by $\hat{\ve}$ and compare our approach with a variational energy-based PINN model. Furthermore, we extend our approach to operator learning, comparing it with a variational physics-informed deep operator network capable of providing solutions for any $\hat{\ve}$ sampled from a domain of interest.

\begin{figure}[t]
    \centering
    \includegraphics[width=0.8\textwidth]{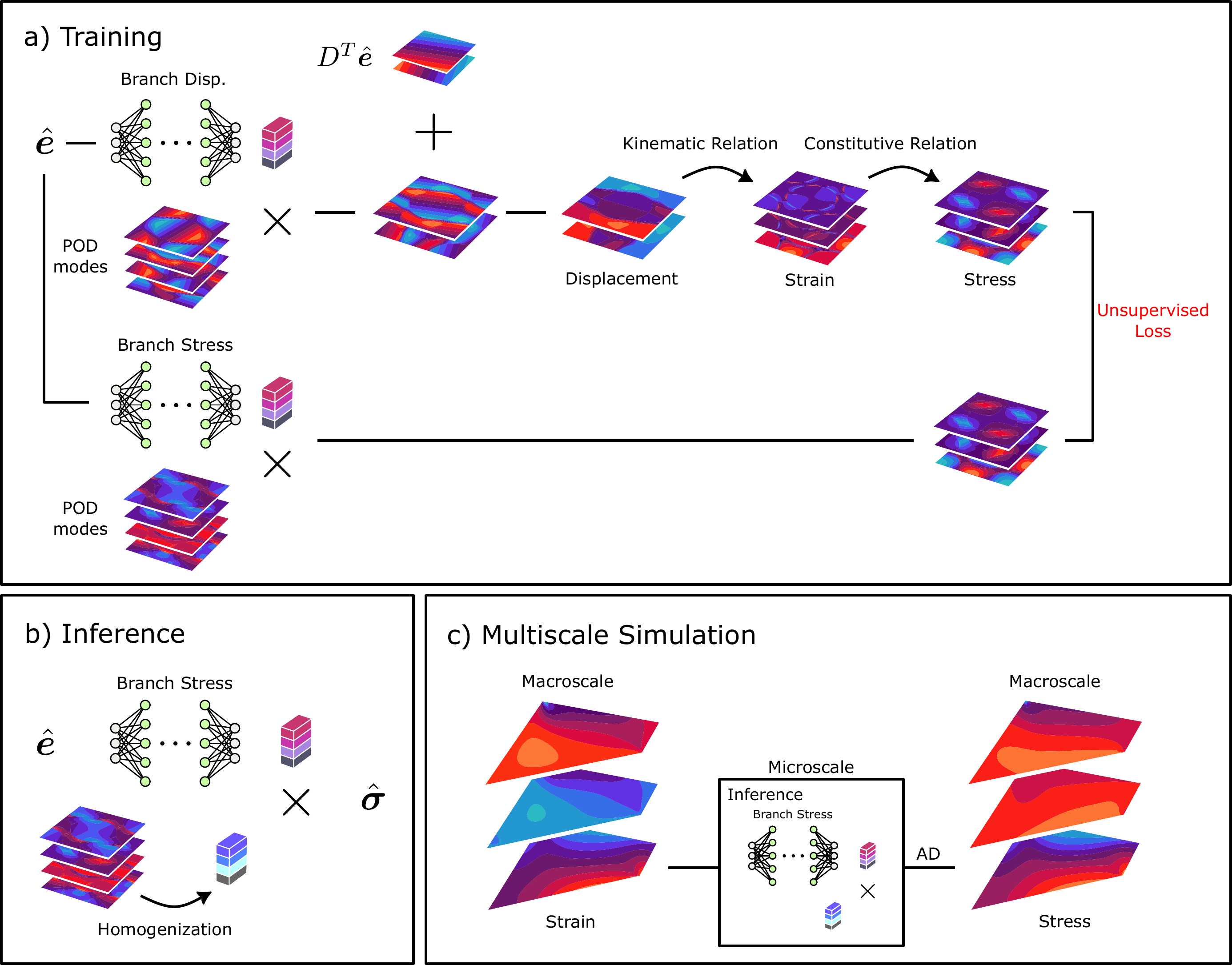}
    \caption{A schematic view of the EquiNO architecture.}
    \label{fig:algo}
\end{figure}

We develop our physics-informed operator network by extending the proposed PINN architecture for inhomogeneous microstructures as described by \citet{HENKES2022114790} to operator networks. \Cref{fig:algo} illustrates a schematic view of the proposed approach.
Consider a limited dataset $\mathcal{D} = \{\vu^i(\vxi), \vsigma^i(\vzeta)\}_{i=1}^{N}$, where $\vu^i(\vxi)$ and $\vsigma^i(\vzeta)$ represent the displacement vector field and the stress tensor field at a spatial discretization of the microstructure comprising nodal points $\vxi$ and integration points $\vzeta$, respectively. The periodic displacement vector field $\ovu^{i}(\vxi)$ in the domain $\Omega$ is obtained by subtracting the prescribed displacement vector field $\mD(\vxi)^{T} \hat{\ve}^i$ from $\vu^i(\vxi)$. We apply POD to $\{\ovu^{i}(\vxi)\}_{i=1}^{N}$ and $\{\vsigma^i(\vzeta)\}_{i=1}^{N}$ separately according to \cref{sec:pod}, leading to two sets of truncated POD modes $\mPhi_{\ovu}$ and $\mPhi_{\vsigma}$,
\begin{equation}
    \mPhi_{\ovu} = \begin{bmatrix}
        \mPhi_{\ovu_1} &
        \ldots &
        \mPhi_{\ovu_d}
    \end{bmatrix}^T \text{ and} \quad 
    \mPhi_{\vsigma} = \begin{bmatrix}
        \mPhi_{\vsigma_{11}} &
        \mPhi_{\vsigma_{12}} &
        \ldots &
        \mPhi_{\vsigma_{dd}}
    \end{bmatrix}^T.
\end{equation}
Note that the snapshot matrices of the vector and tensor fields are constructed such that one set of coefficients is shared among all the vector or tensor components 
\begin{subequations}
    \begin{equation}
        \ovu_i(\vx, \hat{\ve}) = \mPhi_{\ovu_i}^T(\vx)\, \vb_{\ovu}(\hat{\ve}) \quad i=1,\ldots,d,
    \end{equation}
    \begin{equation}
        \vu_i(\vx, \hat{\ve}) = \ovu_i(\vx, \hat{\ve}) + \mD(\vx)^T \hat{\ve},
    \end{equation}
    \begin{equation}
        \vsigma_{ij}(\vx, \hat{\ve}) = \mPhi_{\vsigma_{ij}}^T(\vx)\, \vb_{\vsigma}(\hat{\ve}) \quad i,j=1,\ldots,d.
    \end{equation}
\end{subequations}
By substituting this reduced-order representation into the kinematic relation, \cref{eq:kinematic}, the strain tensor field takes the form  
\begin{equation}
    \vepsilon(\vx) = \mPhi_{\vepsilon}^T(\vx)\, \vb_{\ovu}(\hat{\ve}) + \hat{\ve}, \text{ where} \quad 
    \mPhi_{\vepsilon}(\vx) = \frac{1}{2} \left( \nabla \mPhi_{\ovu}(\vx) + \nabla^T \mPhi_{\ovu}(\vx) \right).
\end{equation}
Projecting the balance of linear momentum for an elastic material in the absence of body forces, \cref{eq:balanceMom}, onto the POD modes and using the linearity of differentiation leads to  
\begin{equation}
    (\nabla \cdot \mPhi_{\vsigma}^T(\vx))\, \vb_{\vsigma}(\hat{\ve}) = \bm{0}.
\end{equation}
Since the POD modes $\mPhi_{\vsigma}(\vx)$ are computed from data that inherently satisfy the balance of linear momentum, they are divergence-free up to numerical discretization errors, i.e., $\nabla \cdot \mPhi_{\vsigma}(\vx) = \bm{0}$, and therefore the governing equation holds independently of the coefficients $\vb_{\vsigma}(\hat{\ve})$. This ensures that the reduced-order model preserves equilibrium by construction and without requiring additional constraints. We refer to these POD modes as divergence-free modes. Furthermore, the periodic boundary condition is also satisfied by construction. Since the displacement modes $\mPhi_{\ovu}(\vx)$ inherently conserve periodicity, the reduced-order model ensures compatibility with the boundary conditions, \cref{eq:bc}, independent of $\vb_{\ovu}(\hat{\ve})$. The aforementioned structural properties, which satisfy both the boundary conditions and the governing equations, are the keys to the superior performance of the proposed method.

Our goal is to learn the solution operator of the microstructure for different boundary conditions imposed by the prescribed global strain $\hat{\ve}$. To achieve this, we consider two branch networks, such that
\begin{subequations}
    \begin{equation}
        \vb_{\ovu}(\hat{\ve}) \approx \gN_{\ovu}(\hat{\ve}; \vtheta_{\ovu}) \in \sR^{p},
    \end{equation}
    \begin{equation}
        \vb_{\vsigma}(\hat{\ve}) \approx \gN_{\vsigma}(\hat{\ve}; \vtheta_{\vsigma}) \in \sR^{p},
    \end{equation}
\end{subequations}
where $p$ indicates the number of POD modes, which are selected to be equal for both networks; however, they may also be configured differently as required. Considering the constitutive relation $\bm{h}$, two sets of stress tensor fields are computed from an input $\hat{\ve}$
\begin{subequations}
    \begin{equation}
        \tilde{\vsigma}(\vx, \hat{\ve}; \vtheta_{\ovu}) = \bm{h} (\mPhi_{\vepsilon}^T(\vx)\, \gN_{\ovu}(\hat{\ve}; \vtheta_{\ovu}) + \hat{\ve}),
    \end{equation}
    \begin{equation}
        \tilde{\vsigma}(\vx, \hat{\ve}; \vtheta_{\vsigma}) = \mPhi_{\vsigma}^T(\vx)\, \gN_{\vsigma}(\hat{\ve}; \vtheta_{\vsigma}),
    \end{equation}
\end{subequations}
where $\tilde{\vsigma}(\vx, \hat{\ve}; \vtheta_{\ovu})$ strongly satisfies the boundary conditions, the kinematic relation, and the constitutive relation, but not the linear balance of momentum. In contrast, $\tilde{\vsigma}(\vx, \hat{\ve}; \vtheta_{\vsigma})$ strongly satisfies the linear balance of momentum. By minimizing the discrepancy between $\tilde{\vsigma}(\vx, \hat{\ve}; \vtheta_{\ovu})$ and $\tilde{\vsigma}(\vx, \hat{\ve}; \vtheta_{\vsigma})$, we derive the solution through a physics-informed, unsupervised learning approach that ensures compliance with all underlying physical constraints. The loss for training neural networks $\gN_{\ovu}$ and $\gN_{\vsigma}$ is obtained according to 
\begin{equation}
    \label{eq:loss}
\gL(\vtheta_{\ovu}, \vtheta_{\vsigma}) = \dfrac{1}{N_f~m~d}\sum_{i=1}^{N_f} \sum_{j=1}^{m} \sum_{k=1}^{d} \left | \tilde{\vsigma}_k(\vx_j, \hat{\ve}_i^f; \vtheta_{\ovu}) - \tilde{\vsigma}_k(\vx_j, \hat{\ve}_i^f; \vtheta_{\vsigma}) \right |^2.
\end{equation}
In this equation, $N_f$ denotes the number of unsupervised inputs $\hat{\ve}^f$ sampled from the domain of interest, where the superscript $f$ indicates that there is no reference data available for these samples, and the networks learn solely from physical principles. The variable $m$ represents the number of collocation points, and $d$ is the number of components of the stress tensor. Imposing physical principles as hard constraints results in a singular term in the loss function, in contrast with existing physics-informed machine-learning frameworks that involve multiple loss terms. This approach potentially simplifies the optimization process by avoiding conflicting gradients. In summary, the EquiNO model effectively incorporates physical properties, ensuring that:
\begin{itemize}
    \item The balance of linear momentum is satisfied by construction.
    \item Periodic boundary conditions are satisfied by construction.
    \item Kinematic and constitutive relations are directly employed during model training.
\end{itemize}
The proposed method  is agnostic to specific architectural constraints, which allows the architecture of the neural networks $\gN_{\ovu}$ and $\gN_{\vsigma}$ to be tailored to the unique demands of various problems. For instance, in the context of time-dependent materials, FNOs may be employed, providing the flexibility required for varying temporal resolutions of the input.

\subsection{Variational physics-informed models}

Variational PINNs (VPINNs) \citep{Kharazmi_vpinn} represent an evolution of traditional PINNs by incorporating the weak (variational) form of PDEs into the neural network framework. By utilizing integration by parts, VPINNs lower the order of derivatives required, making them suitable for problems with sharp gradients. The deep energy method (DEM), as introduced by \citet{Samaniego_dem}, takes advantage of the variational structure of certain BVPs. In DEM, the system's total potential energy is formulated and used as the loss function for training the DNN. A fundamental aspect of this method is the approximation of the body's energy using a weighted sum of the energy density evaluated at integration points.

Inspired by the variational principles underlying VPINNs and DEM, we propose a method for learning the solution operators of RVEs without relying on paired datasets. We consider the weak form of the PDE as the loss function within a discretized domain. The FE discretization allows for analytical differentiation using the shape functions over each domain element, eliminating the dependence on computationally intensive automatic differentiation algorithms. By dividing the domain into elements and employing high-order polynomial approximations within each, our method enhances accuracy and convergence rates. This methodology shares significant similarities with the recent work by \citet{Eshaghi_vino}, particularly in its approach to operator learning and the integration of variational principles.

We develop our variational physics-informed operator network based on DeepONet architecture \citep{lu2021learning}. Let us consider a branch network $\gN_b$ and a trunk network $\gN_t$. We obtain 
\begin{equation}
    \vu(\vx, \hat{\ve}) \approx \tilde{\vu}(\vx, \hat{\ve}) = \gN_t(\vx, m_{\vx}; \vtheta_t)^T\, \gN_b(\hat{\ve};\vtheta_b) + \mD(\vx)^T \hat{\ve},
\end{equation}
where $\vtheta_b$ and $\vtheta_t$ indicate the weights and biases of the branch and trunk networks, respectively, and $m_{\vx}$ is the material index for point $\vx$.  Note that the basis functions for the displacement components are computed by the trunk network and share one set of coefficients. Further, we utilize differentiation of finite element shape functions and the constitutive relations to predict strain and stress tensor fields over the discretized RVE domain. 
The energy of the RVE is approximated by a weighted sum of the elastic strain energy evaluated at the integration points and is used as the physics-informed loss function,
\begin{equation}
    \gL(\vtheta_{b}, \vtheta_{t}) 
    = \dfrac{1}{2\,N_f}\sum_{i=1}^{N_f} \sum_{j=1}^{m} \sum_{k=1}^{d} 
    \tilde{\vsigma}_k(\vx_j, \hat{\ve}^f_i)\,
    \tilde{\vepsilon}_k(\vx_j, \hat{\ve}^f_i)\,
    \omega_j.
    \label{eq:dem}
\end{equation}
Here, $\omega_j$ denotes the quadrature weight associated with the integration point $\vx_j$. Standard Gauss--Legendre quadrature is employed within each finite element. Periodic boundary conditions are enforced as a hard constraint by configuring the trunk network to predict basis functions that are periodic by construction, following the approach of \citet{Dong2021}.

In addition to the variational physics loss, we optionally incorporate a supervised data loss to further guide the training process. The total loss is formulated as a weighted combination of the physics-informed and data-driven objectives, where the weighting is determined adaptively using the multi-task learning strategy proposed by \citet{multi-task-learning}. When included, the data loss is defined as the mean-squared error between the predicted and reference displacement and stress fields evaluated on a limited set of labeled microscale solutions. In the remainder of the article, we refer to this formulation as the variational physics-informed deep operator network (VIONet).

\subsection{Training}

The training process is conducted in an unsupervised manner for EquiNO and physics-informed models. We utilize Latin Hypercube Sampling (LHS) \citep{LHS} to generate $N_f$ unsupervised instances of the global prescribed strain $\hat{\ve}^f$ for training. Optionally, $N_s$ labeled samples are incorporated for supervised training in addition to unsupervised physics-informed training. The network parameters are optimized using the Adam \citep{adam} and limited-memory Broyden--Fletcher--Goldfarb--Shanno (L-BFGS) \citep{liu1989} algorithms, following common practice in training PINNs. We initiate the training process with the Adam optimizer, using a learning rate of $1\times10^{-3}$, and continue it for 1,000 epochs. Subsequently, we switch to the L-BFGS algorithm and allow the training process to continue until convergence. When training with the Adam optimizer, we employ a batch-size training approach to better exploit the stochasticity of the optimizer and avoid falling into local minima. For L-BFGS, we use full-batch training.

\subsection{Homogenization}

In linear homogenization using FEM, the goal is to derive the macroscopic or effective properties of a heterogeneous material. The homogenized or macroscopic stress $\hat{\vsigma}$ is computed as the volume integral of the microscopic stress $\vsigma$ over the domain $\Omega$
\begin{equation}
    \hat{\vsigma} = \frac{1}{|\Omega|} \int_{\Omega} \vsigma(\vx, \hat{\ve}) \, d\Omega,
\end{equation}
where $|\Omega|$ is the volume of the RVE domain $\Omega$. This integral essentially provides an averaged value that represents the overall stress behavior across the material. Considering the FE discretization, the integral can be computed as a volume-weighted average
\begin{equation}
    \hat{\vsigma} = \frac{1}{|\Omega|} \sum_{i=1}^{m} \omega_i\, \vsigma(\vx_i, \hat{\ve}),
    \label{eq:homogenization}
\end{equation}
where $\omega_i$ is the quadrature weight corresponding to the integration point $\vx_i$. In the case of EquiNO, the projection of \cref{eq:homogenization} onto the POD modes yields
\begin{subequations}
    \begin{equation}
        \hat{\mPhi}_{\vsigma} = \frac{1}{|\Omega|}  \sum_{i=1}^{m} \omega_i\, \mPhi_{\vsigma}(\vx_i),
    \end{equation}
    \begin{equation}
        \hat{\vsigma} = \hat{\mPhi}_{\vsigma}^T\, \vb_{\vsigma}(\hat{\ve}) \approx \hat{\mPhi}_{\vsigma}^T\, \gN_{\vsigma}(\hat{\ve}; \vtheta_{\vsigma}),
        \label{eq:Ghomogenization}
    \end{equation}
\end{subequations}
where $\hat{\mPhi}_{\vsigma}$ represents the homogenized POD modes of stress, which can be computed offline. \Cref{eq:Ghomogenization} indicates that the computation of homogenized stresses can be performed in a reduced-order form. This approach eliminates the need to calculate the full stress field and offers significant computational efficiency. Furthermore, the consistent tangent matrix $\hat{\mC}$ is obtained from \cref{eq:Ghomogenization} using automatic differentiation.

\section{Results and discussion}
\label{sec:results-and-discussion}

In this section, we evaluate the performance of the physics-informed operator-learning methods discussed in \cref{sec:methodology} for the simulation of various RVEs and compare them with state-of-the-art methods. Furthermore, we conduct a series of multiscale simulations where the microscale level is modeled using the operator networks. We compare the results with the reference \FEsq in terms of accuracy and computational efficiency. 

\subsection{Problem setup}

We consider a matrix material modeled with a nonlinear elastic material behavior originally extracted from a viscoplastic constitutive relation \citep{hartmannpom04}. The fibers or inclusions are assumed to behave linearly, in an isotropic elastic manner, in which the material law takes the form
\begin{equation}
    \label{eq:linElas}
    \vsigma(\vx) = K_f \tr(\vepsilon(\vx))\sI + 2G_f \vepsilon^{\textrm{D}}(\vx), \quad \vx \in \Omega^f,
\end{equation}
with scalar values of bulk modulus $K_f$ and shear modulus $G_f$, where $\Omega^f \subset \Omega$ represents the domain occupied by the fiber material. The nonlinear constitutive relation for the matrix material reads 
\begin{equation}
    \label{eq:nonlinElas}
    \vsigma(\vx) = K_m \tr(\vepsilon(\vx))\sI + G_m(\vepsilon^{\textrm{D}}(\vx))\,\vepsilon^{\textrm{D}}(\vx), \quad \vx \in \Omega^m,
\end{equation}
for any point $\vx$ in the domain occupied by the matrix material $\Omega^m \subset \Omega$ with the deformation-dependent shear modulus
\begin{equation}
    G_m(\vepsilon^{\textrm{D}}(\vx)) = \frac{\alpha_1}{\alpha_2 + \vert\vert\bm{\varepsilon}^{\textrm{D}}(\vx)\vert\vert_2}.
\end{equation}
The nonlinear elastic material is characterized by the bulk modulus $K_m$ and the scalar parameters $\alpha_1$ and $\alpha_2$. In \cref{eq:linElas,eq:nonlinElas}, $\sI$ represents the second-order identity tensor, and $\bm{\varepsilon}^{\textrm{D}}$ refers to the deviatoric part of the strain tensor. The corresponding values for these material parameters are provided in \cref{tab:matparRVE}.
\begin{table}[ht]
    \caption{Material parameters for the elastic fiber and the nonlinear elastic matrix within the RVEs.}
    \label{tab:matparRVE}
    \centering
    \begin{threeparttable}
    \begin{tabular}{c c c c c}
    \toprule
    $K_f$ & $G_f$ & $K_m$ & $\alpha_1$ & $\alpha_2$ \\
    \si{\N\per\mm\squared} & \si{\N\per\mm\squared} & \si{\N\per\mm\squared} & \si{\N\per\mm\squared} & - \\
    \midrule
    \num{4.35e+04} & \num{2.99e+04} & \num{4.78e+03} & \num{5.0e+01} & \num{6.0e-02} \\
    \bottomrule
    \end{tabular}
    \end{threeparttable}
\end{table}

\subsection{Microscale simulation}

\begin{figure}[ht]
    \centering
        \includegraphics[width=0.1\textwidth]{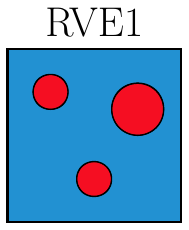}
        ~~~~\includegraphics[width=0.1\textwidth]{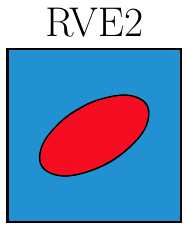}
        ~~~~\includegraphics[width=0.1\textwidth]{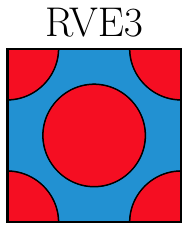}
    \caption{RVEs considered in this study. The blue and red regions indicate nonlinear and linear elastic materials, respectively.}
    \label{fig:rves}
\end{figure}

The performance of EquiNO and VIONet in the simulation of three two-dimensional microscale RVEs, illustrated in \cref{fig:rves}, is investigated. 
We utilize 100 samples to compute the POD modes for the EquiNO models, whereas the training of the VIONet models does not require any data. 
We select multilayer perceptrons (MLPs) with four hidden layers and 64 neurons in each hidden layer as the architecture of the neural networks. The branch and trunk networks are equipped with \textit{swish} and \textit{tanh} activation functions, respectively. An ablation study on the influence of network size and number of POD modes is provided in \cref{app:ablation}. The effect of the number of snapshots used for construction of the POD modes is analyzed in a later section. The models are implemented and trained using the TensorFlow \citep{tensor_flow} and JAX \citep{jax2018github} frameworks.

\subsubsection{EquiNO vs. VPINN}

\begin{figure}[ht]
    \centering
    \begin{subfigure}{0.8\textwidth}
        \centering
        \includegraphics[width=0.35\textwidth]{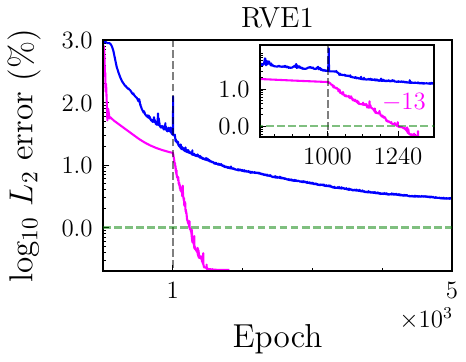}
        \hfill
        \includegraphics[width=0.6\textwidth]{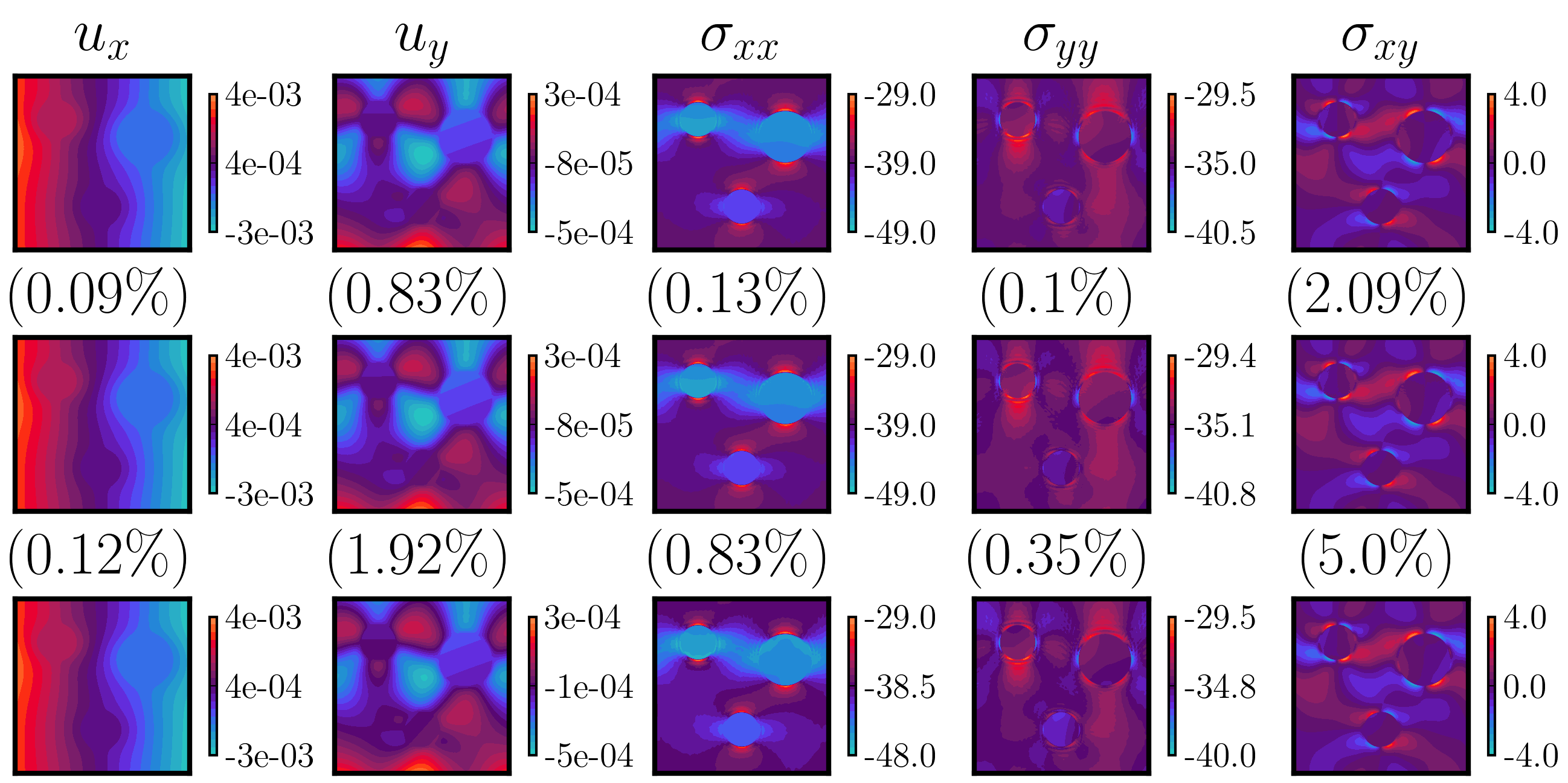}
        \caption{}
    \end{subfigure}

    \begin{subfigure}{0.8\textwidth}
        \centering
        \includegraphics[width=0.35\textwidth]{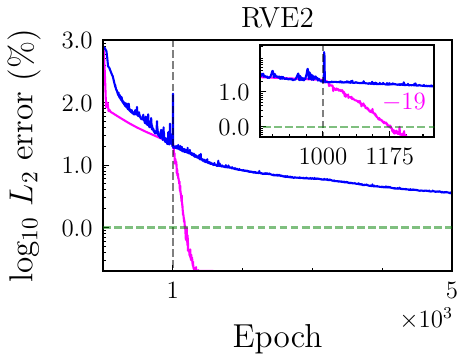}
        \hfill
        \includegraphics[width=0.6\textwidth]{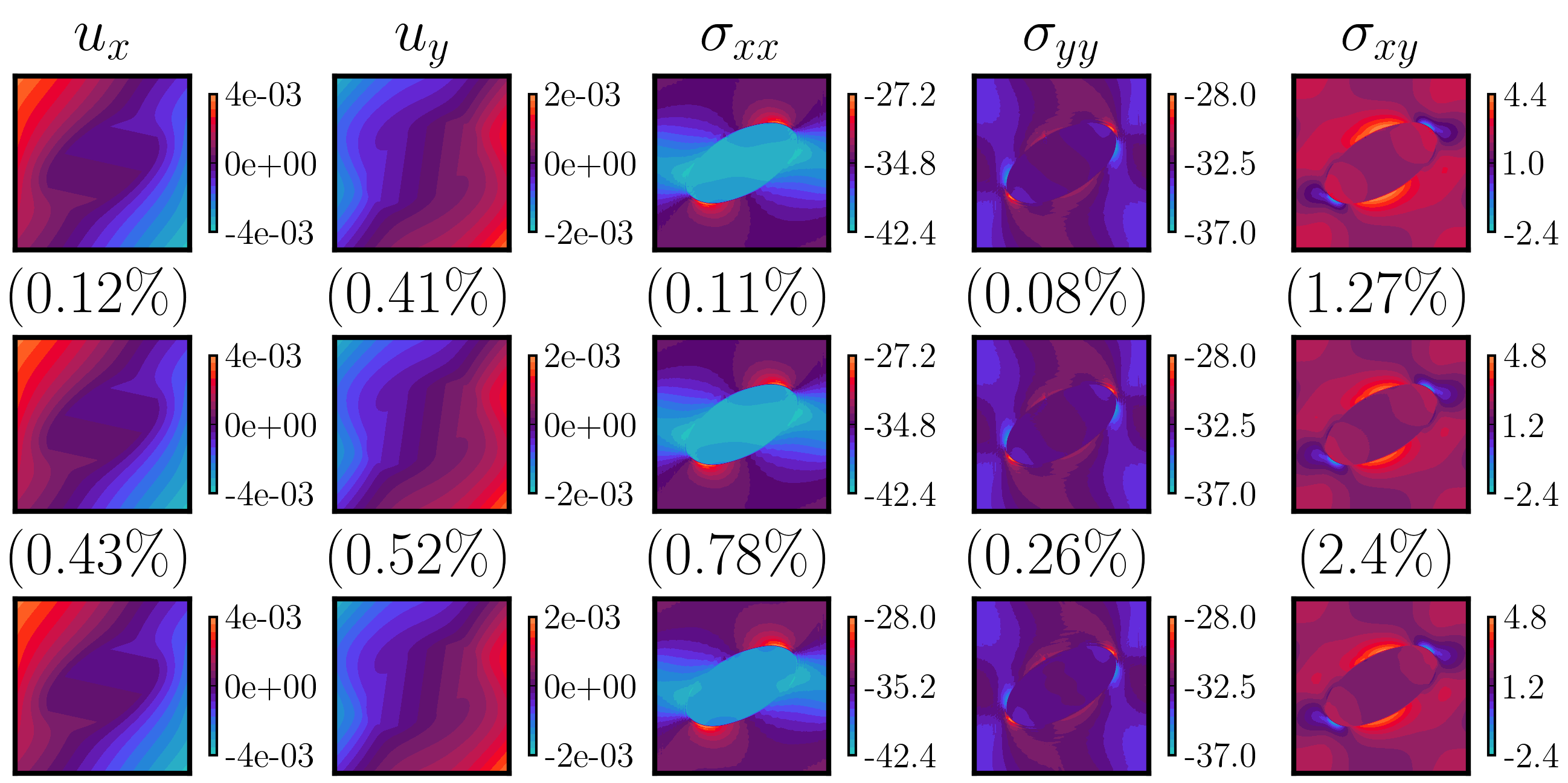}
        \caption{}
    \end{subfigure}

    \begin{subfigure}{0.8\textwidth}
        \centering
        \includegraphics[width=0.35\textwidth]{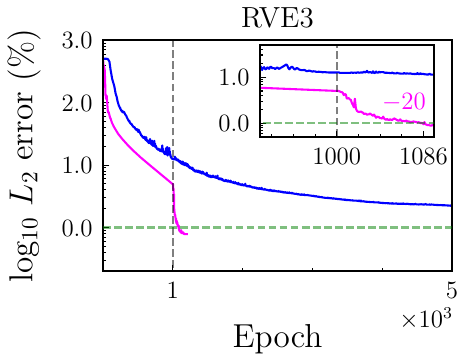}
        \hfill
        \includegraphics[width=0.6\textwidth]{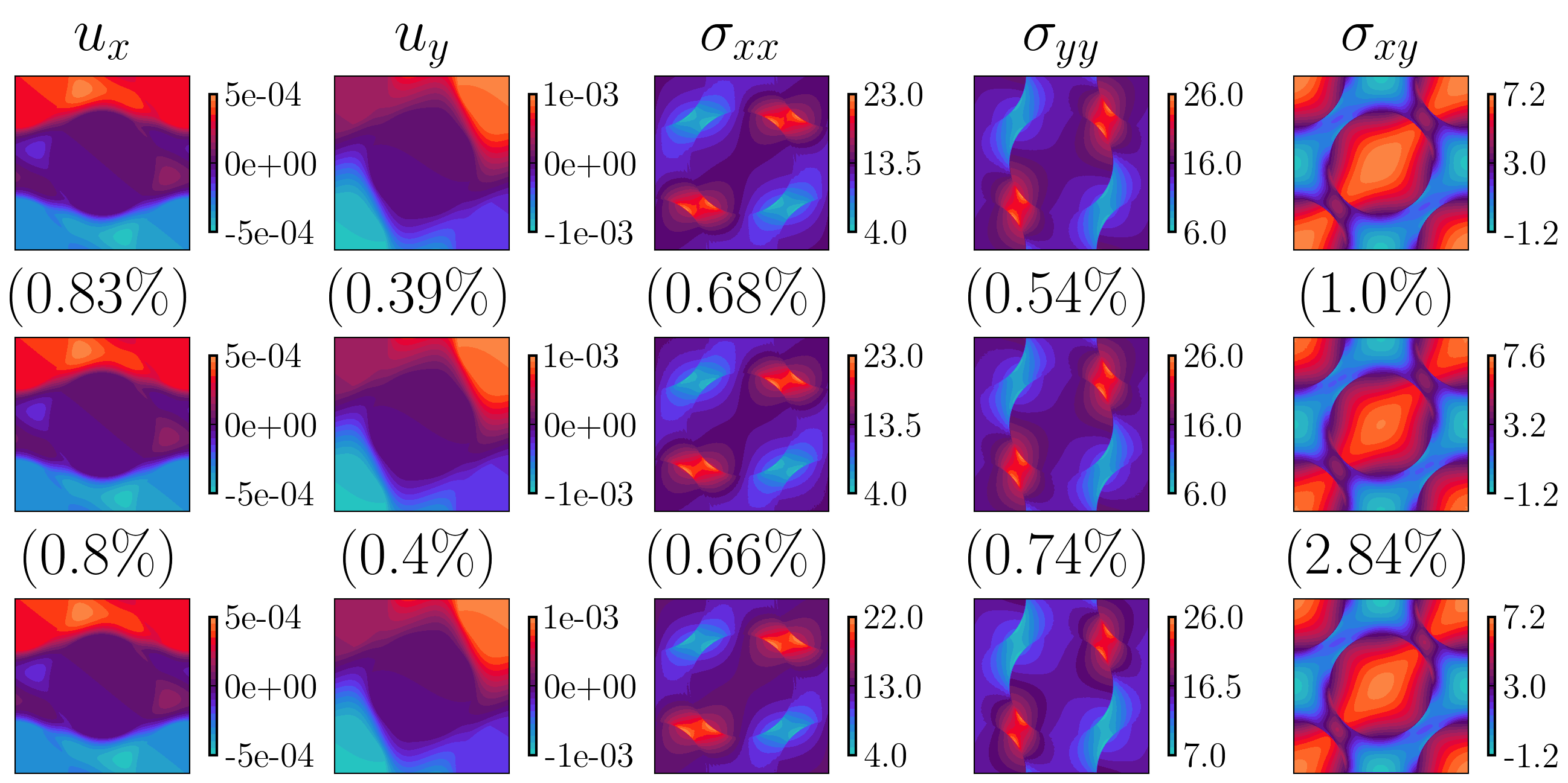}
        \caption{}
    \end{subfigure}
    
        \caption{Comparison of the convergence behavior of EquiNO (magenta) and VPINN (blue) for three different RVEs. The plots on the left show the relative $L_2$ stress error percentages during training, averaged over ten independent samples and all stress components. The field visualizations on the right display the reference (top) and the predicted displacement and stress components for EquiNO (middle) and VPINN (bottom), with the relative $L_2$ error percentages indicated in parentheses.}

    \label{fig:pinns}
    \end{figure}

We first employ EquiNO and a variational PINN model to solve the governing equations of RVEs for one prescribed global strain, $\hat{\ve}$. The objective of this comparison is to investigate the convergence rate and accuracy of the models. To this end, we utilize a VPINN model that is similar to VIONet but is specifically designed for one $\hat{\ve}$ and consists of only one neural network, $\gN$, which obtains the solution as
\begin{equation}
    \vu(\vx, \hat{\ve}) \approx \tilde{\vu}(\vx, \hat{\ve}) = \gN(\vx; \vtheta)  + \mD(\vx)^T \hat{\ve}.
\end{equation}
The loss function for this VPINN model is the same as in \cref{eq:dem}, with $N_f = 1$. To compare the convergence rate of EquiNO and VPINN, we take ten independent samples for $\hat{\ve}$ and train a separate model for each. Our experiments are conducted for all three RVEs to ensure a comprehensive evaluation of the models across different microstructural configurations. The relative $L_2$-norm of the errors in stress during training is illustrated in \cref{fig:pinns}. The plots depict the averaged values across all ten samples and three stress components. The error curves show a clear distinction between the two approaches in terms of convergence behavior. From the results, we observe that EquiNO demonstrates significantly faster convergence compared to VPINN, especially after the transition to the L-BFGS optimization algorithm. This accelerated learning process can be attributed to the efficient utilization of the 100 precomputed samples used to generate the POD modes, and to defining the loss directly over the stress values. EquiNO achieves rapid adaptation to new, unseen boundary conditions. In contrast, VPINN exhibits a slower convergence pattern, requiring a greater number of epochs to reach a comparable error level.
The visualized displacement and stress fields on the right in \cref{fig:pinns} show that both models can provide accurate solutions after training. However, EquiNO produces stress distributions with lower relative errors, whereas VPINN tends to exhibit slightly larger deviations, particularly in complex stress components such as $\sigma_{xy}$.

\subsubsection{EquiNO for physics-informed operator learning}

We continue by comparing EquiNO with a range of physics-informed (PI) and data-driven (DD) operator-learning models on RVE1. EquiNO and VIONet were introduced in $\S$\ref{sec:methodology}. VINO, proposed by \citet{Eshaghi_vino}, and VIOFormer (variational physics-informed operator transformer) baselines follow the same variational training strategy as VIONet but differ in the choice of operator-learning backbone. VINO replaces the DeepONet backbone with an FNO \citep{li2020fourier}, while VIOFormer employs an operator transformer (OFormer) \citep{li2023transformer}. In both cases, the loss function is identical to that of VIONet and is computed by minimizing the elastic strain energy given in \cref{eq:dem}, thereby isolating the effect of the operator architecture while keeping the physics-informed training procedure fixed. Architectural details of the FNO and OFormer baselines are provided in \cref{app:net}.

\begin{figure}[ht]
    \centering
    \includegraphics[width=0.8\textwidth]{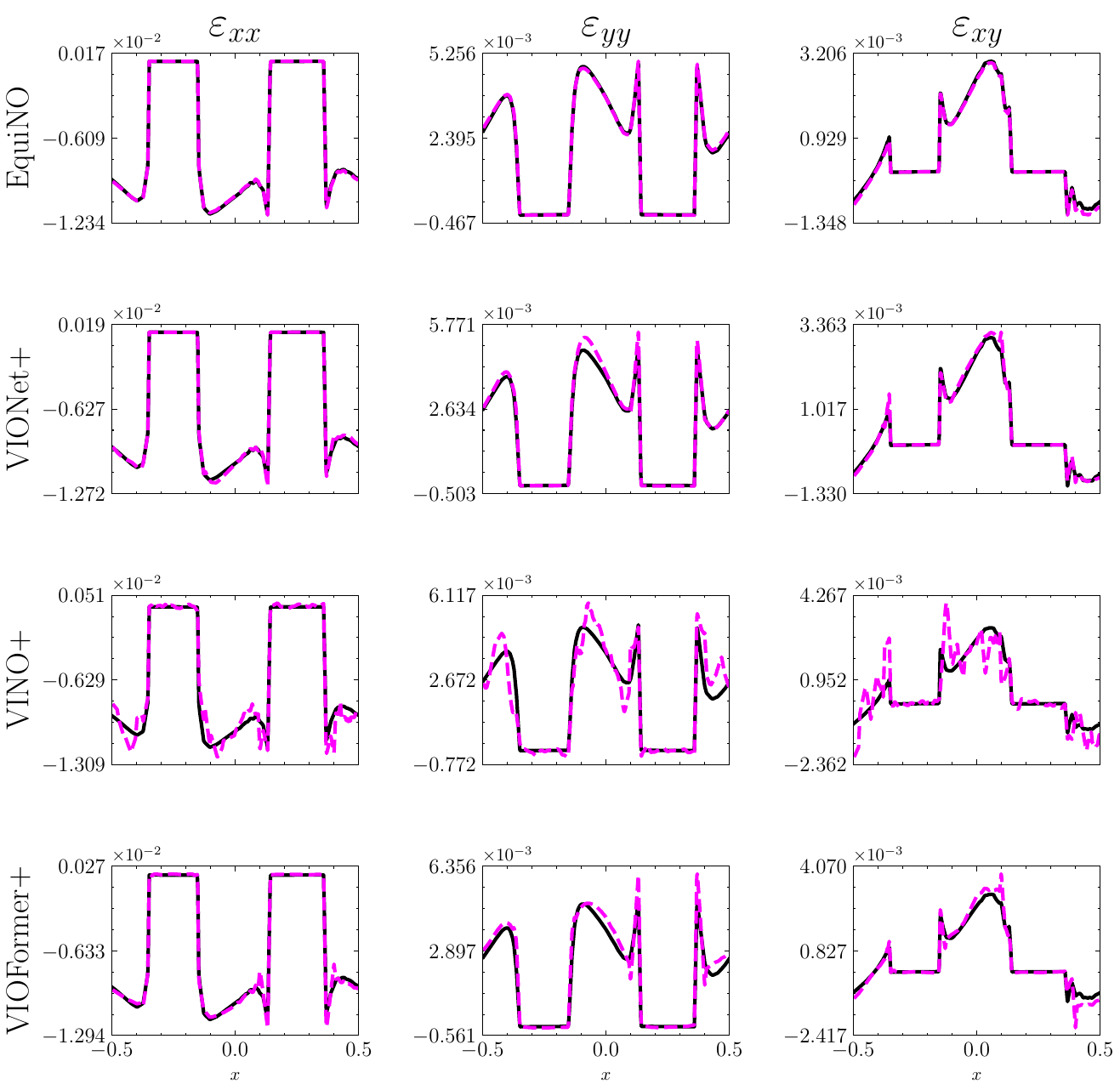}
    \caption{Reference (solid black line) and predicted (dashed magenta line) strain components along the line $y=0.25$ for the RVE1 obtained from different physics-informed operator-learning models.}
    \label{fig:lineplot_comp_ons}
\end{figure}

For a fair comparison, all models indicated by a ``+'' include a supervised loss component in their training objective derived from the same set of $N_s = 100$ microscale solution fields, which are used either for constructing reduced bases or for supervised training, depending on the model. Physical laws are enforced weakly for VIONet, VINO, and VIOFormer, while EquiNO enforces equilibrium by construction through divergence-free POD bases. All PI models are trained using $N_f = 2000$ unsupervised input samples generated via LHS and evaluated on a test set of 100 samples. In contrast, DD models are trained purely in a supervised manner using the same full-field data and do not explicitly enforce physical constraints.
\begin{table}[ht]
    \centering
    \resizebox{\textwidth}{!}{%
    \begin{threeparttable}  
    \caption{Relative $L_2$-norm of errors on a test dataset containing 100 samples for each component of displacement vector field and stress tensor field. $\mu_{\vsigma}$ shows the mean relative $L_2$ error of the predicted stress components. Predictions are obtained using different physics-informed (PI) and data-driven (DD) operator-learning models for RVE1. Column $N_{\rm{param}}$ indicates the number of parameters in the model and column $T_{\rm{inf}}$ represents the relative inference time with respect to the EquiNO model.} 
    \label{tab:NOs1}
    \small
    \begin{tabular}{@{}lccccccccc@{}}
    \toprule
    \textbf{Models} &  Training  & $u_x(\%)$ & $u_y(\%)$ & $\sigma_{xx}(\%)$ & $\sigma_{yy}(\%)$ & $\sigma_{xy}(\%)$ & $\mu_{\vsigma}(\%)$ & $N_{\rm{param}}$ & $T_{\rm{inf}}$\\
    \midrule
    EquiNO & PI & 1.1 & 2.01 & 2.35 & 1.51 & 6.82 & \textbf{3.56} & 27K & 1.0\\
    VIONet+ & PI & 1.54 & 1.61 & 5.79 & 4.75 & 7.37 & 5.97 & 37K & 1.5\\ 
    VINO+ \citep{Eshaghi_vino} & PI & 8.6 & 18.54 & $>$100 & $>$100 & $>$100 & $>$100 & 199K & 6.0\\
    VIOFormer+ & PI & 3.3 & 3.25 & 25.66 & 23.16 & 104.03 & 50.95 & 30K & 20.0\\
   \midrule
   DeepONet \citep{lu2021learning} & DD &  2.71 & 2.4 & 7.07 & 5.47 & 10.4 & 7.65 & 37K & 1.5\\
   FNO \citep{li2020fourier} & DD &  27.58 & 37.09 & $>$100 & $>$100 & $>$100 & $>$100
    & 199K & 6.0\\
   OFormer \citep{li2023transformer} & DD & 12.67 & 17.56 & 85.59 & 68.18 & 60.63 & 71.47 & 30K & 20.0\\
    \bottomrule
    \end{tabular}
    \end{threeparttable}
    }
\end{table}
\Cref{tab:NOs1} and \cref{fig:lineplot_comp_ons} provide a detailed comparison of the predictive performance of different operator-learning models on RVE1. Among all physics-informed approaches, EquiNO achieves the lowest errors for stress components with the mean relative $L_2$-norm error of $\mu_{\vsigma} = 3.56\%$, despite having the smallest number of trainable parameters and the lowest inference cost. The strain profiles in \cref{fig:lineplot_comp_ons} further confirm that EquiNO accurately captures sharp variations induced by material heterogeneities, with predictions closely matching the reference solution across all components. VIONet+ represents the strongest variational baseline and yields competitive displacement accuracy. However, noticeable deviations appear in the predicted strain and stress components, particularly in regions with pronounced gradients. These discrepancies are reflected in the higher stress errors reported in \cref{tab:NOs1}.
In contrast, VINO+ and VIOFormer+ exhibit severe degradation in stress prediction, with errors exceeding 100\% for several components in the case of VINO+. As illustrated in \cref{fig:lineplot_comp_ons}, these models fail to accurately resolve sharp transitions in the strain fields, which leads to oscillatory behavior and pronounced error amplification when computing stresses. For models based on FNOs, this behavior can be attributed to the combination of global spectral representations, which are known to suffer in the presence of sharp gradients or discontinuities \citep{lu2019deeponet}, and the high sensitivity of stress computation to small displacement errors in heterogeneous media. 

A similar trend is observed for the purely data-driven models. DeepONet exhibits inferior stress accuracy compared to physics-informed approaches, with a relative error of 10.4\% for the shear component, while FNO and OFormer suffer from catastrophic failures in stress prediction. Finally, it is worth noting that attention-based architectures, such as VIOFormer and OFormer, incur substantially higher inference costs, as they process all grid points simultaneously and rely on dense dot-product attention, which scales quadratically with the number of spatial degrees of freedom.
\begin{figure}[bht]
    \centering
    \includegraphics[width=0.35\textwidth]{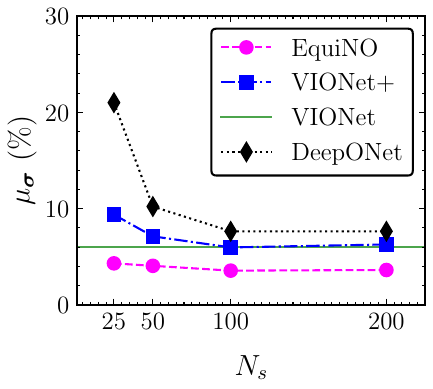}
    \caption{Mean relative $L_2$ error of the predicted stress components, $\mu_{\vsigma}$, as a function of the number of supervised training samples.}
    \label{fig:data_efficiency}
\end{figure}

To further assess data efficiency, we compare EquiNO with VIONet and the fully data-driven DeepONet as a function of the number of supervised training samples. \Cref{fig:data_efficiency} reports the mean relative $L_2$ error of the predicted stress components, $\mu_{\vsigma}$, for increasing amounts of labeled data. In the case of EquiNO, the data are solely used to construct POD basis functions. It can be observed that EquiNO exhibits consistently low errors. In contrast, VIONet+ benefits from additional supervision but saturates at a higher error level, while DeepONet shows pronounced data dependency and poor performance in low-data regimes. These results highlight the superior data efficiency of EquiNO. A slight performance degradation is observed for VIONet+ at very low data regimes compared to the purely physics-informed model, VIONet. This behavior can be attributed to the introduction of a data loss that competes with the physics residual, leading to suboptimal optimization.

\subsubsection{Hard vs. soft enforcement of physics in operator learning}

Next, we employ the proposed physics-informed neural operator to learn the solution operator of RVEs, as discussed in \cref{sec:methodology}. As before, the models are trained using $N_f = 2000$ unsupervised input samples generated via LHS and tested on a dataset containing 100 samples. In this context, \emph{hard enforcement} refers to restricting the hypothesis space such that equilibrium is satisfied by construction, as in EquiNO, whereas \emph{soft enforcement} imposes physical laws through penalty terms in the loss function, as in VIONet. The results for all three RVEs are summarized in \cref{tab:NOs}, where the performance of VIONet and VIONet+ is compared against that of EquiNO. The results highlight the comparative performance of the EquiNO and VIONet models across different components of the displacement vector field and stress tensor field. The EquiNO model consistently demonstrates superior performance in terms of lower relative $L_2$-norm errors for each stress component and across all RVEs. For instance, for RVE3, EquiNO achieved significantly lower errors in both displacement components and stress components compared to VIONet, resulting in a mean error $\mu_{\vsigma}$ of 3.69\% versus 9.65\% for VIONet. Similar trends are observed in RVE2 and RVE3, where EquiNO maintains its advantage, particularly in stress prediction. 

The results illustrate the advantages of enforcing equilibrium as a hard constraint. By preserving equilibrium by construction, EquiNO yields improved stability and consistently low stress errors across all RVEs. Overall, EquiNO achieves mean stress errors below 5\% for all considered microstructures. An ablation study isolating the effect of the divergence-free constraint is provided in \cref{app:dfm}.

\begin{table}[ht]
    \centering
    \begin{threeparttable}  
    \caption{Relative $L_2$-norm of errors on a test dataset containing 100 samples for each component of displacement vector field and stress tensor field. Predictions are obtained using the EquiNO and VIONet models for all three RVEs.}
    \label{tab:NOs}
    \small
    \begin{tabular}{@{}lccccccc@{}}
    \toprule
    \textbf{Models} & RVE & $u_x(\%)$ & $u_y(\%)$ & $\sigma_{xx}(\%)$ & $\sigma_{yy}(\%)$ & $\sigma_{xy}(\%)$ & $\mu_{\vsigma}(\%)$\\
    \midrule
    EquiNO & 1 & 1.1 & 2.01 & 2.35 & 1.51 & 6.82 & \textbf{3.56} \\
    VIONet+ & 1 &1.54 & 1.61 & 5.79 & 4.75 & 7.37 & 5.97 \\
    VIONet & 1 & 0.99 & 0.89 & 4.13 & 3.10 & 10.94 & 6.06 \\
    \midrule
    EquiNO & 2 & 1.77 & 1.53 & 2.89 & 2.61 & 7.68 & \textbf{4.39} \\
    VIONet+ & 2 & 2.75 & 2.24 & 6.41 & 6.87 & 11.03 & 8.1 \\
    VIONet & 2 & 1.22 & 1.22 & 5.00 & 5.26 & 15.74 & 8.67 \\ 
    \midrule
    EquiNO & 3 & 1.11 & 1.18 & 2.37 & 2.2 & 6.51 & \textbf{3.69} \\
    VIONet+ & 3 & 2.69 & 2.35 & 5.78 & 5.87 & 8.59 & 6.75  \\
    VIONet & 3 & 2.56 & 2.35 & 5.31 & 5.62 & 18.01 & 9.65 \\
    \bottomrule
    \end{tabular}
    \end{threeparttable}  
\end{table}

\begin{figure}[ht]
    \centering
    \includegraphics[width=0.8\textwidth]{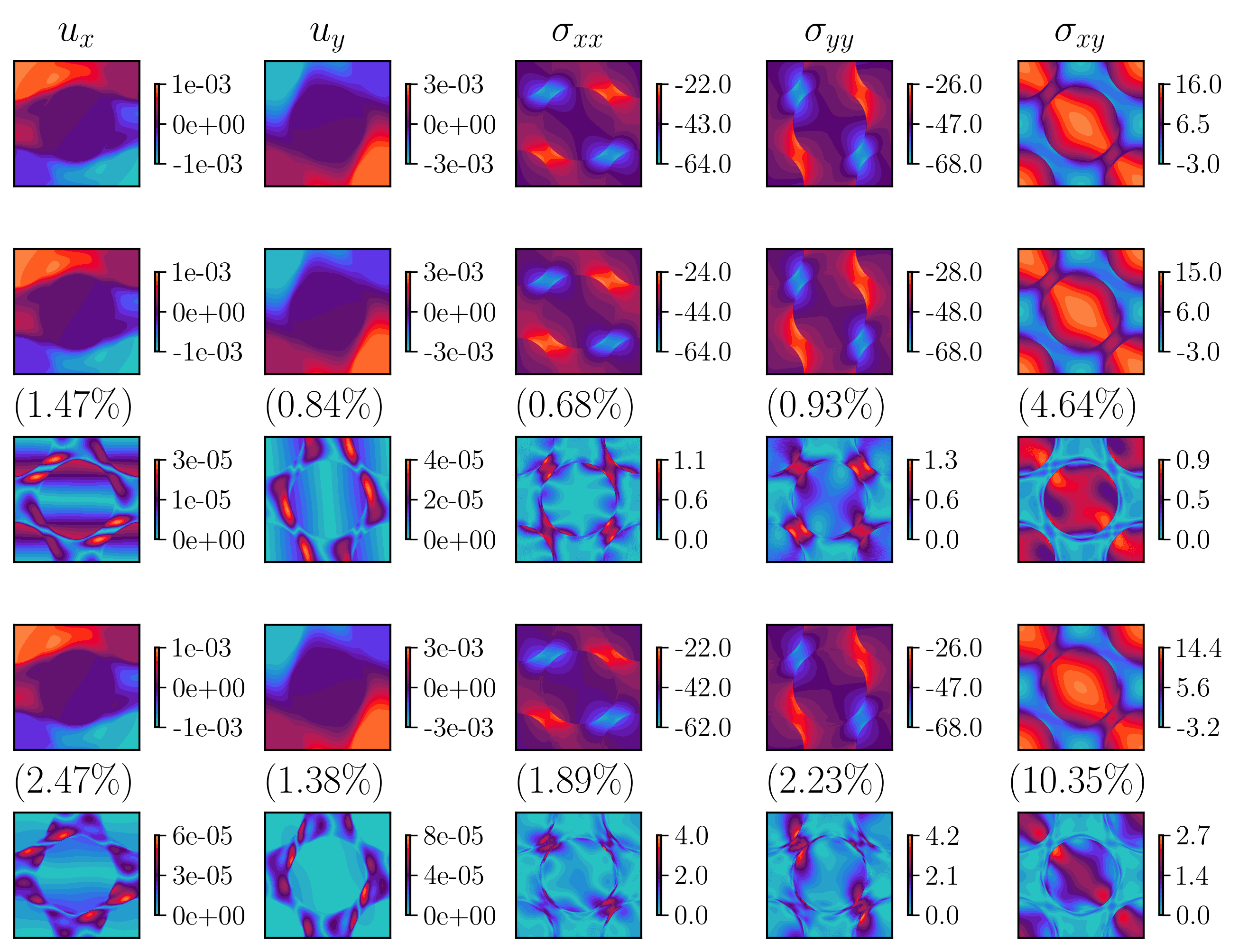}
    \caption{RVE3 field visualizations of the reference (first row) and the predicted displacement and stress components for EquiNO (second row) and VIONet (fourth row). Results correspond to a sample that exhibits the median relative $L_2$ error on stress components among the test samples. The third and fifth rows display the absolute errors in the predictions obtained from EquiNO and VIONet, respectively. The relative $L_2$ errors are reported in parentheses.}
    \label{fig:NOs-rve1}
\end{figure}

A field visualization of the reference data and predictions is provided in \cref{fig:NOs-rve1} for RVE3. This visualization corresponds to a sample that exhibits the median relative $L_2$ error on stress components among the test samples. 
The figure illustrates a comparison between the reference displacement and stress fields (first row) and the predicted fields from the EquiNO (second row) and VIONet (fourth row) models. Both predictions demonstrate a strong match with the reference fields, as reflected by low relative $L_2$ errors, such as 1.47\% and 0.84\% for the normal stresses obtained from EquiNO. However, VIONet shows higher errors, particularly notable in the shear stress component at 10.35\%. The third and fifth rows highlight the absolute errors in each model's predictions, underscoring EquiNO's superior accuracy across the components. The higher errors in the shear component can be attributed to the nonlinear material properties, including the deformation-dependent shear modulus, which creates additional challenges for accurate predictions. Furthermore, the scale difference between the normal and shear stress components, nearly an order of magnitude, can bias the learning process towards the normal components when the loss function is formulated as the strain energy or when it averages the discrepancies among the components. EquiNO demonstrates a robust capability to mitigate these issues and deliver more accurate solutions. This is evident in the third and fifth rows, which display the absolute errors in each model's predictions, emphasizing EquiNO's superior performance across components.

Further, the trained physics-informed operator networks are utilized to simulate the microstructure in the context of a multiscale simulation, where the macroscale is simulated using the FE method. We refer to this approach as FE-OL in the remainder of the article.

\subsubsection{Three-dimensional microstructure}

\begin{figure}[ht]
    \centering
        \begin{subfigure}{0.25\textwidth}
             \includegraphics[width=\textwidth]{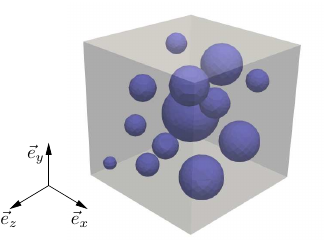}
             \caption{}
             \label{fig:rve3d:a}
        \end{subfigure}
        \hfil
        \begin{subfigure}{0.25\textwidth}
             \includegraphics[width=\textwidth]{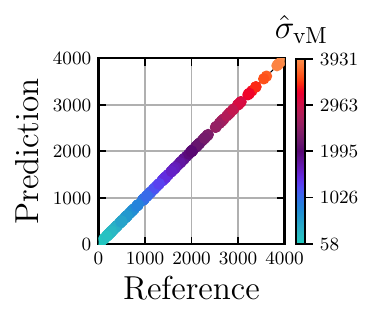}
             \caption{}
             \label{fig:rve3d:b}
        \end{subfigure}
    \hfill
        \begin{subfigure}{\textwidth}
        \includegraphics[width=\textwidth]{con_rve_3d.png}
        \caption{}
        \label{fig:rve3d:c}
        \end{subfigure}
    \caption{(a) Geometry of the three-dimensional RVE. Gray and blue regions denote nonlinear and linear elastic materials, respectively; (b) comparison between reference and predicted homogenized von Mises stress responses; (c) three-dimensional visualizations of the stress components, showing the reference solution (top) and the corresponding EquiNO prediction (bottom) for a test sample with median relative $L_2$ error among all stress components. The relative $L_2$ errors for each component are reported in parentheses.}
    \label{fig:rve3d}
\end{figure}

Next, we employ EquiNO for physics-informed surrogate modeling of a three-dimensional RVE, illustrated in \cref{fig:rve3d:a}, to demonstrate the scalability of the proposed method. The material parameters are identical to those used in the two-dimensional RVEs and are reported in \cref{tab:matparRVE}. A total of 100 precomputed microscale solutions are used to construct the POD bases. In this setting, the resulting POD basis functions are three-dimensional, while the branch networks predict their corresponding coefficients. Consequently, the architectures of the branch networks $\gN_{\ovu}(\hat{\ve}; \vtheta_{\ovu})$ and $\gN_{\vsigma}(\hat{\ve}; \vtheta_{\vsigma})$ remain unchanged compared to the two-dimensional cases and are not directly affected by the spatial dimensionality of the problem. As in the two-dimensional studies, we employ multilayer perceptrons with four hidden layers and 64 neurons per layer. The model is trained using $N_f = 2000$ unsupervised input samples generated via LHS and evaluated on a test set containing 200 samples. The results are summarized in \cref{fig:rve3d}.

The three-dimensional results further confirm the accuracy of the proposed EquiNO framework. As shown in \cref{fig:rve3d:b}, the predicted homogenized von Mises stress closely matches the reference solution over the full range of loading conditions. Moreover, the field-level comparisons in \cref{fig:rve3d:c} demonstrate excellent agreement between the predicted and reference stress distributions for all tensor components, with relative $L_2$ errors below $1\%$ for this representative sample. The mean relative $L_2$ error on the test data across all stress components is $0.22\%$. These results highlight EquiNO's ability to accurately capture complex three-dimensional stress patterns in heterogeneous microstructures while strictly satisfying equilibrium by construction. Regarding computational performance, the inference time for predicting the full three-dimensional displacement and stress fields is approximately $0.35$ seconds per batch of 100 samples, compared to about $0.15$ seconds for the two-dimensional RVEs. This increase is primarily due to the higher cost of reconstructing three-dimensional fields from the POD expansion. In contrast, the inference time for predicting homogenized stress quantities remains comparable to the two-dimensional case. This is because the three-dimensional POD modes are first homogenized into a low-dimensional $p$-vector, after which the prediction pipeline is identical to that used in the two-dimensional setting.

\subsection{Multiscale simulation}

\begin{figure}
    \centering
    \hfill
    \begin{subfigure}{0.3\textwidth}
        \centering
        \begin{tikzpicture}
            \node[anchor=south west,inner sep=0] (image) at (0,-0.1) 
                {\includegraphics[height=0.7\textwidth]{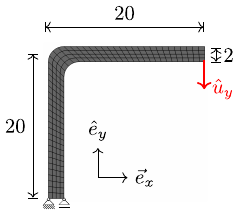}};
        \end{tikzpicture}
        \caption{L-profile.}
        \label{fig:subplot1}
    \end{subfigure}
    \hfill
    \begin{subfigure}{0.3\textwidth}
        \centering
        \begin{tikzpicture}
            \node[anchor=south west,inner sep=0] (image) at (0,0) 
                {\includegraphics[height=0.7\textwidth]{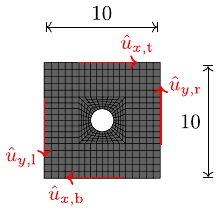}};
        \end{tikzpicture}
        \caption{Plate with hole.}
        \label{fig:subplot2}
    \end{subfigure}
    \hfill
    \begin{subfigure}{0.3\textwidth}
        \centering
        \begin{tikzpicture}
            \node[anchor=south west,inner sep=0] (image) at (0,0) 
                {\includegraphics[height=0.7\textwidth]{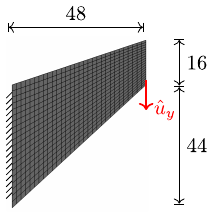}};
        \end{tikzpicture}
        \caption{Cook's membrane.}
        \label{fig:subplot3}
    \end{subfigure}
    \hfill~
    
    \caption{Macroscale test cases employed for testing the performance of the FE-OL method against the reference \FEsq method.}
    \label{fig:macro_tests}
\end{figure}

In this section, we investigate the use of physics-informed operator networks for multiscale simulations in three macroscale test cases: L-profile, plate with a hole, and Cook's membrane, illustrated in \cref{fig:macro_tests}. We employ 8-noded quadrilateral elements for spatial discretization. Consequently, each macroscale element contains $n_{g} = 9$ integration points. As a result, this setup necessitates $n_{e} \times n_{g}$ calls to the RVE in every global Newton iteration of a \FEsq computation. For the L-profile, the spatial discretization consists of $n_{e} = 200$ elements and $n_{n} = 709$ nodes. A displacement boundary condition is prescribed on the top right edge, given by $\hat{u}_y(t) = \SI{-3}{\mm\per\s}\,t$. In the plate example, load functions are defined as follows: $\hat{u}_{x,\text{top}}(t) = \hat{u}_{y,\text{right}}(t) = \SI{4e-02}{\mm\per\s}\,t$ and $\hat{u}_{x,\text{bottom}}(t) = \hat{u}_{y,\text{left}}(t) = -\SI{4e-02}{\mm\per\s}\,t$. This setup comprises $n_{e} = 420$ elements and $n_{n} = 1364$ nodes. In the case of Cook's membrane, $n_{e} = 600$ elements and $n_n = 1901$ nodes are used. The left edge is fixed, while a displacement boundary condition of $\hat{u}_y(t) = \SI{2}{\mm\per\s}\,t$ is applied on the right edge. All the numerical examples use an initial time-step size of $\Delta t_0 = \SI{1e-03}{\second}$, with simulations performed for $t \in [0,1]$. The Backward-Euler method is applied for time discretization, which is here used solely for incremental load application to avoid convergence issues. The time-step size $\Delta t$ is dynamically adjusted according to the number of Newton iterations $N_{\text{iter}}$ and the current time-step size $\Delta t_n$, as discussed in \citet{Eivazi2023}. 

\begin{figure}[ht]
    \centering
    \includegraphics[width=0.8\textwidth]{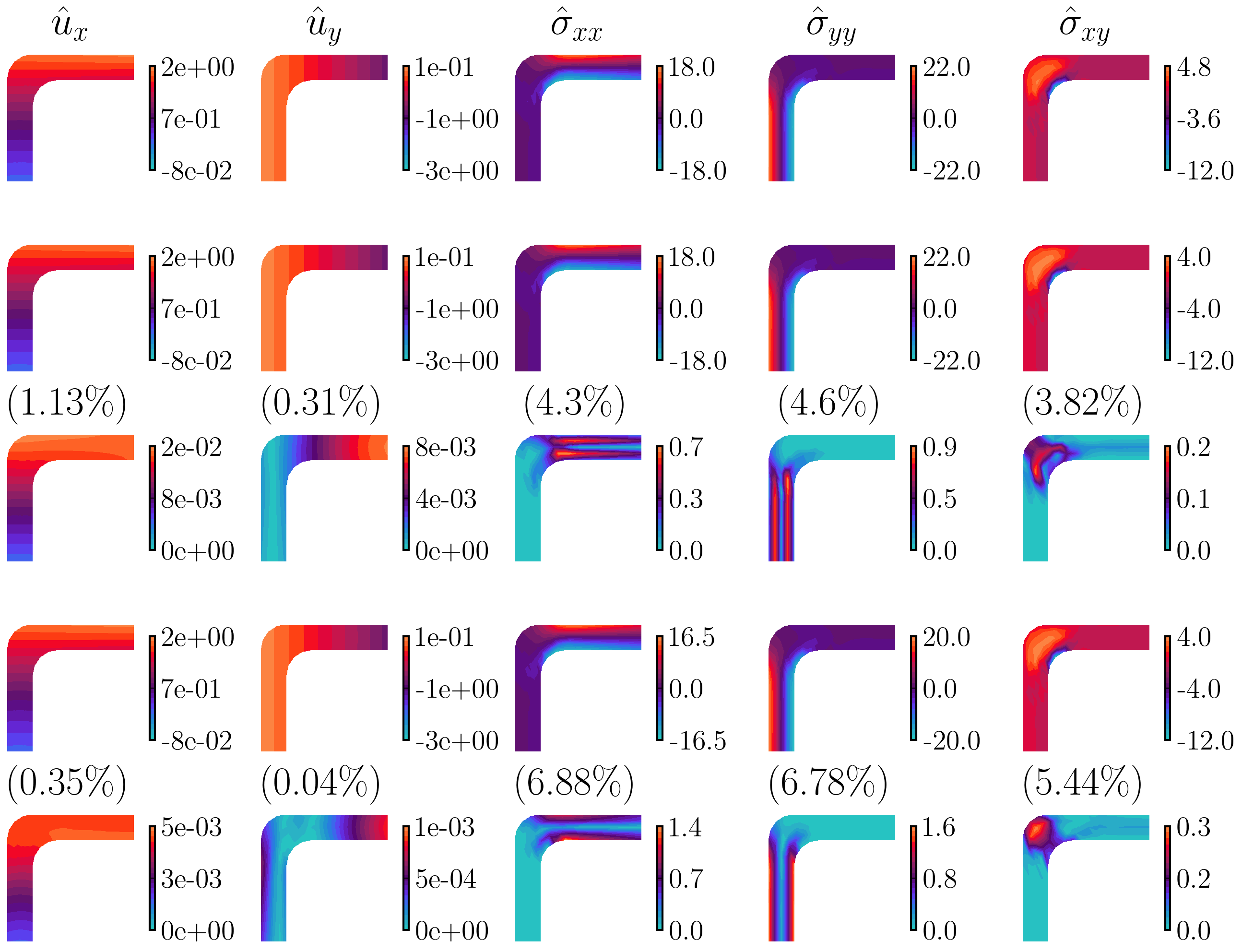}
    \caption{Field visualizations of the reference (first row) and the predicted displacement and stress components obtained from multiscale simulation using EquiNO (second row) and VIONet (fourth row) for the L-profile with RVE3. The third and fifth rows display the absolute errors in the predictions of EquiNO and VIONet, respectively. The relative $L_2$ errors are reported in parentheses. The plots are zoomed in for the range $x = [0, 10]$ and $y = [10, 20]$ to enhance visualization.}
    \label{fig:lprofile_rve3}
\end{figure}

\begin{figure}[ht]
    \centering
    \includegraphics[width=0.8\textwidth]{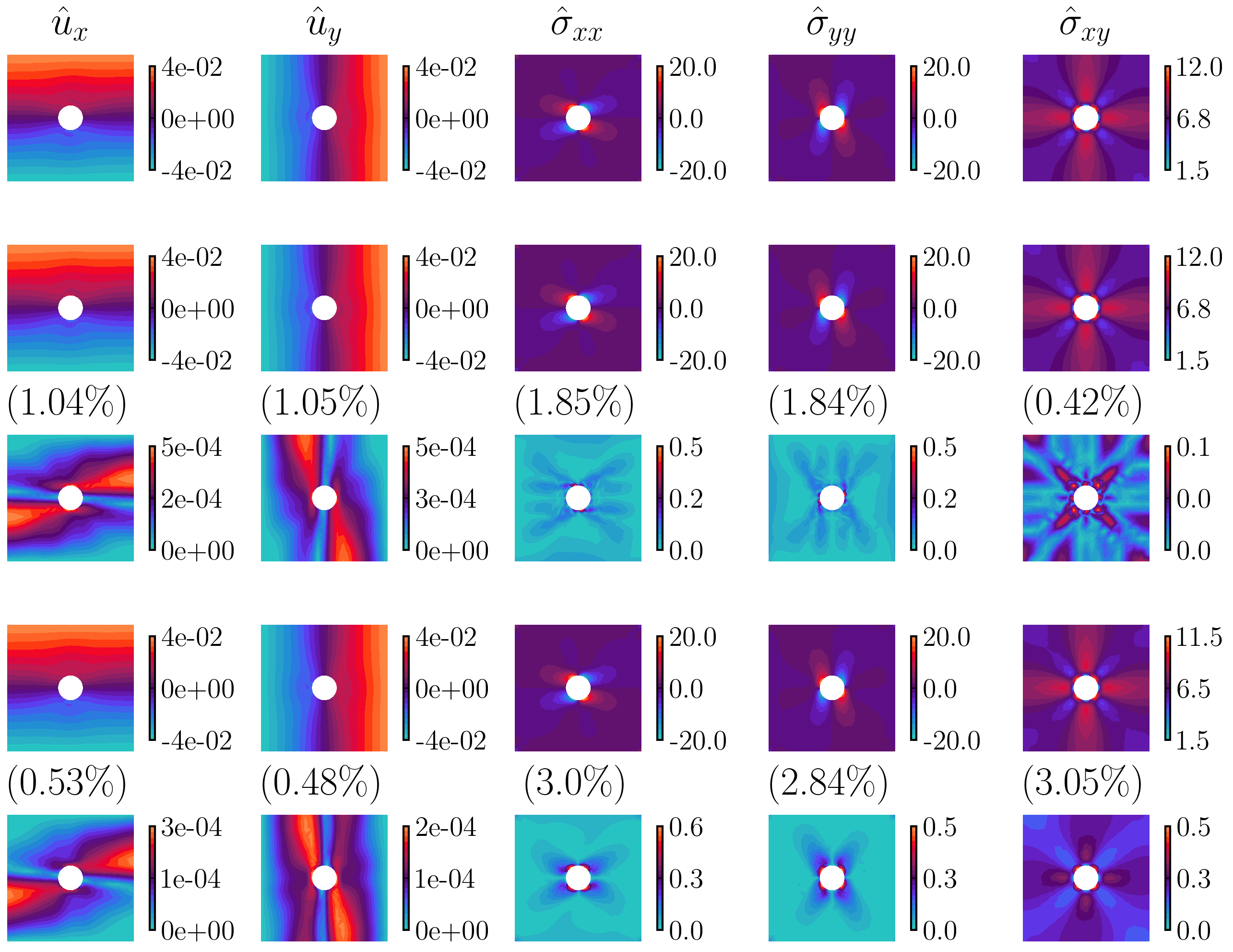}
    \caption{Field visualizations of the reference (first row) and the predicted displacement and stress components obtained from multiscale simulation using EquiNO (second row) and VIONet (fourth row) for the plate with RVE2. The third and fifth rows display the absolute errors in the predictions of EquiNO and VIONet, respectively. The relative $L_2$ errors are reported in parentheses.}
    \label{fig:plate_rve3}
\end{figure}

\begin{figure}[ht]
    \centering
    \includegraphics[width=0.8\textwidth]{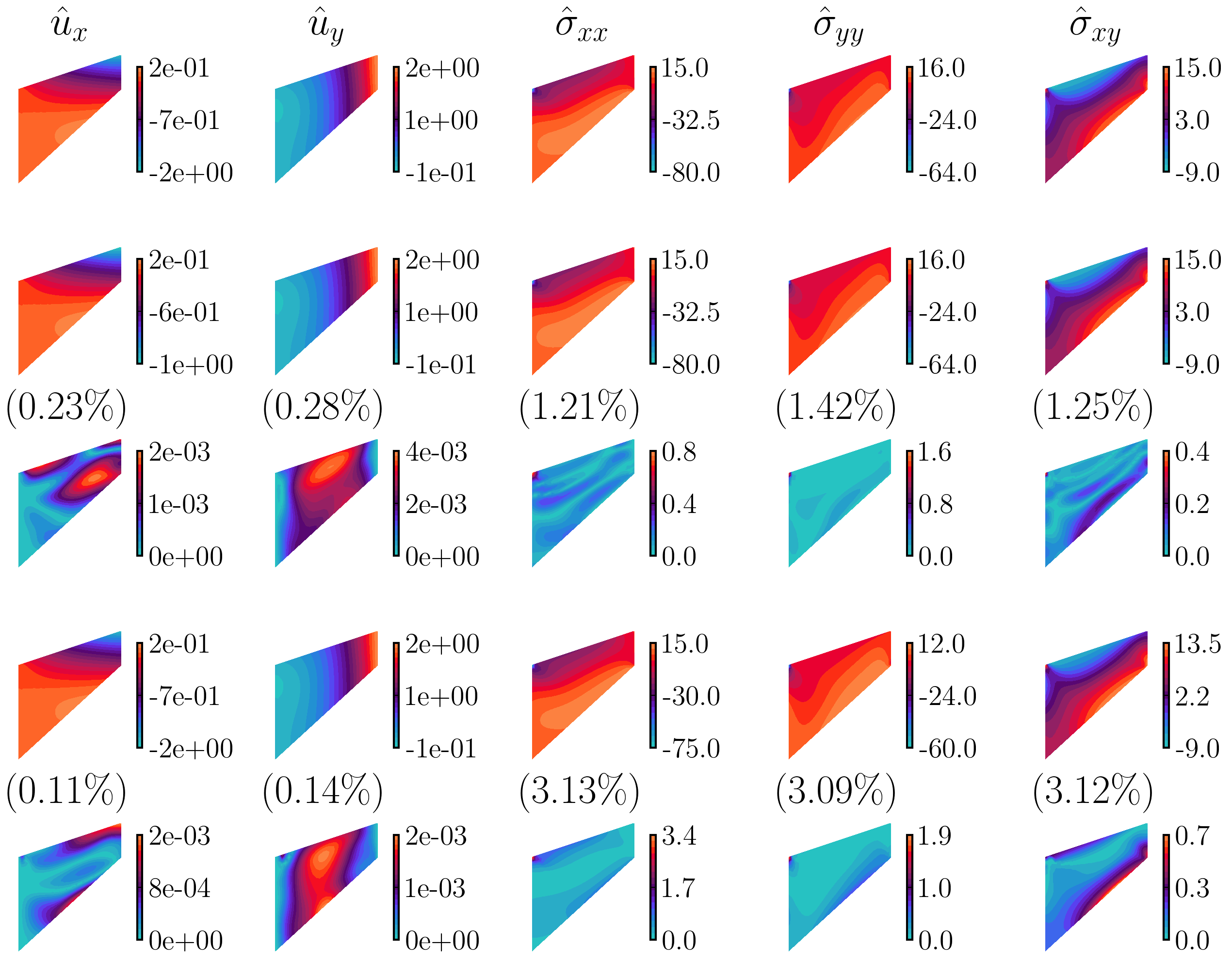}
    \caption{Field visualizations of the reference (first row) and the predicted displacement and stress components obtained from multiscale simulation using EquiNO (second row) and VIONet (fourth row) for the Cook's membrane with RVE1. The third and fifth rows display the absolute errors in the predictions of EquiNO and VIONet, respectively. The relative $L_2$ errors are reported in parentheses.}
    \label{fig:cook_rve3}
\end{figure}

We employ the microstructures illustrated in \cref{fig:rves} and conduct multiscale simulations of the macroscale test cases using the EquiNO and VIONet models as microscale simulators. Results are summarized in \cref{tab:multiscale_sim}, where we report the relative $L_2$-norm of errors for the displacement vector field and the stress tensor field, as well as the speedup gain compared to the reference \FEsq simulations.
\begin{table}[ht]
    \centering
    \resizebox{\textwidth}{!}{%
    \begin{threeparttable}  
    \caption{Relative $L_2$-norm of errors in the FE-OL simulation results using different physics-informed operator networks with respect to the reference \FEsq data.}
    \label{tab:multiscale_sim}
    \scriptsize
    \begin{tabular}{@{}lccccccccc@{}}
    \toprule
    Test & RVE &  Model  & $u_x(\%)$ & $u_y(\%)$ & $\sigma_{xx}(\%)$ & $\sigma_{yy}(\%)$ & $\sigma_{xy}(\%)$ & $\mu_{\vsigma}(\%)$ & Speedup\\
    \midrule
     L-profile & 1 & EquiNO & 0.73 & 0.17 & 3.48 & 3.6 & 3.62 & 3.57 & 1617 \\
     L-profile & 1 & VIONet &  0.01 & 0.01 & 1.69 & 1.7 & 1.67 & \textbf{1.69} & 188 \\
     \midrule
     L-profile & 2 & EquiNO & 0.92 & 0.21 & 4.5 & 4.67 & 4.56 & 4.58 & 2089 \\
     L-profile & 2 & VIONet & 0.04 & 0.02 & 2.69 & 2.7 & 2.64 & \textbf{2.68} & 207 \\
     \midrule
     L-profile & 3 & EquiNO & 1.13 & 0.31 & 4.3 & 4.6 & 3.82 & \textbf{4.24} & 2944 \\
     L-profile & 3 & VIONet &  0.35 & 0.04 & 6.88 & 6.78 & 5.44 & 6.37 & 189 \\
     \midrule
     Plate & 1 & EquiNO & 0.95 & 0.96 & 1.23 & 1.21 & 0.31 & \textbf{0.92} & 2370 \\
     Plate & 1 & VIONet & 0.38 & 0.37 & 2.16 & 2.13 & 2.28 & 2.19 & 187 \\ 
     \midrule
     Plate & 2 & EquiNO & 1.04 & 1.05 & 1.85 & 1.84 & 0.42 & \textbf{1.37} & 4501 \\
     Plate & 2 & VIONet & 0.53 & 0.48 & 3.0 & 2.84 & 3.05 & 2.96 & 278 \\ 
     \midrule
     Plate & 3 & EquiNO & 0.45 & 0.44 & 1.24 & 1.24 & 0.7 & \textbf{1.06} & 5201 \\
     Plate & 3 & VIONet & 7.92 & 7.92 & 7.38 & 7.46 & 6.0 & 6.95 & 198 \\ 
     \midrule
     Cook & 1 & EquiNO & 0.23 & 0.28 & 1.21 & 1.42 & 1.25 & \textbf{1.29} & 4326 \\
     Cook & 1 & VIONet & 0.11 & 0.14 & 3.13 & 3.09 & 3.12 & 3.11 & 245 \\
     \midrule
     Cook & 2 & EquiNO & 0.27 & 0.22 & 1.34 & 1.42 & 1.28 & \textbf{1.35} & 7422 \\
     Cook & 2 & VIONet & 0.32 & 0.19 & 5.05 & 4.85 & 4.96 & 4.95 & 348 \\
     \midrule
     Cook & 3 & EquiNO & 0.18 & 0.08 & 0.98 & 0.99 & 0.89 & \textbf{0.95} & 8253 \\
     Cook & 3 & VIONet & 4.47 & 1.04 & 12.76 & 10.21 & 10.99 & 11.32 & 222 \\
    \bottomrule
    \end{tabular}
    \end{threeparttable}
    }
\end{table}
Multiscale simulation results for the displacement and stress tensor components when utilizing different RVEs as the microstructure are illustrated in \cref{fig:lprofile_rve3,fig:plate_rve3,fig:cook_rve3}. 
For the L-profile test case, both EquiNO and VIONet models deliver accurate results for displacement, with the highest error being 1.13\%. However, errors in the stress components show considerable variation between the two models. The mean relative $L_2$-norm of errors among the stress components, $\mu_{\vsigma}$, is 3.57\%, 4.58\%, and 4.24\% for EquiNO, compared to 1.69\%, 2.68\%, and 6.37\% for VIONet across the three RVEs, respectively. In the plate with a hole test case, EquiNO generally outperforms VIONet in estimating stress components, with $\mu_{\vsigma}$ values of 0.92\%, 1.37\%, and 1.06\% for the three RVEs. VIONet, on the other hand, results in higher errors, specifically 2.19\%, 2.96\%, and 6.95\%. A similar trend is observed in Cook's membrane test case. EquiNO achieves $\mu_{\vsigma}$ values of 1.29\%, 1.35\%, and 0.95\% across the three RVEs, showing its effectiveness in simulating the microscale. Meanwhile, VIONet consistently yields higher errors with values of 3.11\%, 4.95\%, and 11.32\%.

The superior performance of the EquiNO model over the VIONet model is expected, as EquiNO utilizes 100 samples to compute POD modes, providing it with a robust set of basis functions for accurate predictions. VIONet, operating in a completely unsupervised manner, generally provides higher error rates, particularly in stress component estimation. Despite this, VIONet's predictions are still very good, achieving high accuracy in many cases. VIONet's ability to deliver low errors in displacement and reasonable stress predictions for some RVEs demonstrates its potential utility in scenarios where supervised data may not be available. However, VIONet particularly struggles with RVE3 across different test cases, where it consistently shows higher errors in stress components compared to EquiNO. This suggests that while VIONet has promising capabilities, it may face challenges with more complex or specific RVE geometries. Overall, our results emphasize the robust performance of the EquiNO model across various scenarios.

\paragraph{Computational efficiency}
The last column in \cref{tab:multiscale_sim} highlights the speedup factor of multiscale simulations when employing EquiNO or VIONet as surrogate models for simulating the microscale. EquiNO significantly enhances computational efficiency by approximating global stresses $\hat{\vsigma}$ and tangent matrices $\hat{\mC}$ without computing the full state of the RVE as described in \cref{eq:Ghomogenization}. This technique aligns closely with methods used in reduced order modeling. 

In contrast, VIONet requires the full estimation of the RVE before obtaining homogenized quantities. This inherently involves a more computationally intensive process. As a result, EquiNO exhibits performance that is up to an order of magnitude faster than VIONet. For instance, speedup factors in the L-profile tests for EquiNO are 1617, 2089, and 2944, substantially exceeding VIONet's figures of 188, 207, and 189, respectively. Similar advantages are noted in the plate and Cook's membrane tests, where EquiNO consistently achieves speedup factors ranging from 2370 to 8253, compared to VIONet's smaller range of 187 to 348. This efficiency advantage makes EquiNO particularly well-suited for large-scale simulations or many-query scenarios such as topology optimization, where computational resources are a limiting factor. Furthermore, the performance of VIONet could potentially be enhanced by employing the discrete empirical interpolation method (DEIM) \citep{deim}, which would sidestep the computation of the full state of the RVE. However, in this study, our focus is on comparing the speedup gain achieved by operator network models that directly replace RVE simulations. This comparison highlights EquiNO's ability to efficiently streamline computational processes without compromising essential accuracy, showcasing its potential as a powerful tool in multiscale simulation frameworks.
\begin{figure}[ht]
    \centering
    \includegraphics[width=0.3\textwidth]{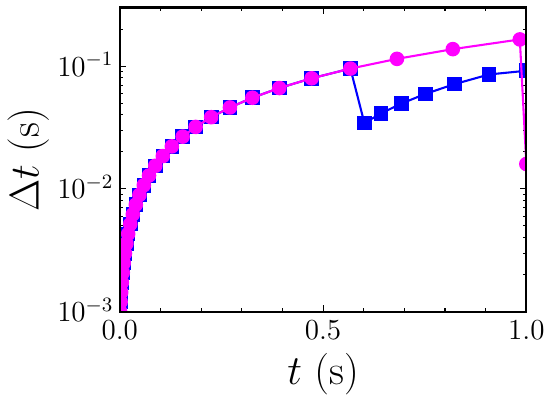}
    \hfill
    \includegraphics[width=0.3\textwidth]{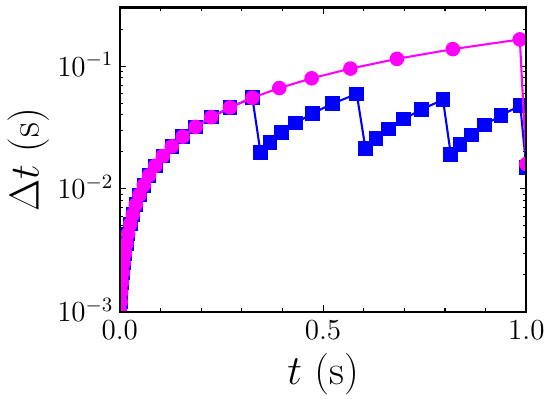}
    \hfill
    \includegraphics[width=0.3\textwidth]{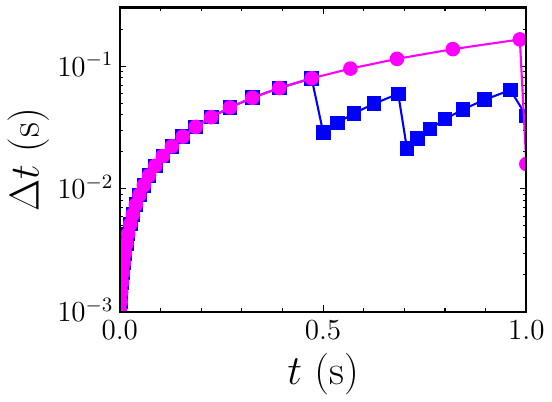}
    \caption{Step-size behavior for multiscale simulation of the plate test case using FE-OL with EquiNO (magenta) against reference \FEsq (blue) for RVE1 to RVE3 from left to right.}
    \label{fig:stepsize}
\end{figure}

The step-size behaviors for the multiscale simulations of the plate test case for three RVEs are shown in \cref{fig:stepsize}. Both the \FEsq and FE-OL with EquiNO methods initially exhibit similar behaviors. However, the \FEsq computation faces a limitation on the time-step size $\Delta t$, which remains below $\SI{1e-01}{\s}$. This constraint results from non-convergence issues in RVE computations for certain load increments. In contrast, the EquiNO model permits a continuously increasing time-step size. This outcome reveals that a surrogate operator network model for microscale evaluations can extend applicable time-step sizes, especially in the non-linear elastic problems considered in this study, leading to a further reduction in computational time.

\section{Summary and conclusions}
\label{sec:conclusions}

 Multiscale problems require solving partial differential equations across multiple scales, presenting significant computational challenges. Techniques addressing these problems are accurate but computationally expensive, making them unsuitable for many-query scenarios such as topology optimization. Recently, deep learning has emerged as a viable approach for surrogate modeling of microscale physics by replacing the microscale with a so-called \textit{substitutive} surrogate model that maps macroscale inputs to outputs while incorporating microscale effects. However, these models often fail to respect specific physical constraints during training and inference, such as the microscale constitutive relations and balance of linear momentum. To address these shortcomings, we introduce a \textit{complementary} surrogate modeling technique using physics-informed operator learning, resulting in the FE-OL framework. This framework simulates microscale behavior using a physics-informed operator network. 
 
 We propose the Equilibrium Neural Operator (EquiNO), an efficient and accurate neural operator specifically designed for multiscale problems in solid mechanics. Our method approximates RVE solutions by projecting governing equations onto a set of POD modes, creating a reduced order model. By exploiting linearity in the balance of momentum and constructing a set of divergence-free stress tensor POD modes, our method inherently preserves equilibrium. The periodic boundary conditions are also satisfied as hard constraints. This approach produces a physics-informed loss function defined on stress components, potentially simplifying optimization by avoiding conflicting gradients. In addition, we implement a variational physics-informed operator network (VIONet) for simulating microscale mechanics, using the weak form of the PDE as the loss function over a discretized domain. Finite element discretization facilitates analytical differentiation through shape functions and eliminates the need for computationally expensive automatic differentiation. Both operator learning approaches are integrated into an efficient MPI-available FE multiscale simulation code written in FORTRAN, forming the FE-OL framework. Advanced techniques and libraries such as JIT, JAX \citep{jax2018github}, and FORPy \citep{forpy} are employed to ensure efficient communication between FORTRAN and Python.

 We initially evaluate the methods by solving a single instance of a microscale boundary value problem. A VPINN model is compared with EquiNO, where, in both cases, the solution is obtained in an unsupervised manner. The results indicate that both methods provide accurate simulations, with EquiNO converging faster due to the utilization of precomputed POD modes. We then use the proposed physics-informed neural operator to learn the solution operator of RVEs, including a three-dimensional case. The resulting predictions are compared with those obtained from state-of-the-art physics-informed and data-driven operator-learning models. All physics-informed models are trained using a set of unsupervised input samples generated via Latin Hypercube Sampling. Results show that EquiNO consistently achieves low relative $L_2$-norm errors for all RVEs, with a maximum mean error ($\mu_{\vsigma}$) of 4.39\% among stress tensor components. It is demonstrated that VIONet also accurately simulates RVEs. However, it presents higher errors, particularly in the shear stress component.

 In multiscale simulations, where the macroscale is evaluated using the FEM, we assess the models' performance on three test cases: L-profile, plate with a hole, and Cook's membrane. EquiNO achieves mean relative $L_2$-norm errors as low as 3.57\% across RVEs, whereas VIONet obtains values up to 11.32\%. EquiNO offers speedup factors exceeding 8000 times that of the reference \FEsq method while maintaining excellent accuracy and requiring a small dataset---100 samples in our experiments. VIONet, trained solely on physical principles, also enhances computational efficiency by two orders of magnitude. However, our findings suggest that physics-informed networks based solely on physics, without data, may require extended training periods, particularly for complex microstructures. Considering the trade-off between data requirements and modeling efficiency, EquiNO emerges as an optimal choice, balancing data-driven and physics-based insights.

 Our study is limited to quasi-static and nonlinear elastic problems in solid mechanics. Accordingly, the proposed methods may be extended to time-dependent materials by utilizing FNOs \citep{li2020fourier} as branch networks to process global strains over time as input functions. Finally, extending the proposed framework to enable generalization across different RVE geometries is an important direction for future work.
 
\section*{Acknowledgments}

Hamidreza Eivazi's research was conducted within the Research Training Group CircularLIB, supported by the Ministry of Science and Culture of Lower Saxony with funds from the program zukunft.niedersachsen of the Volkswagen Foundation (MWK$|$ZN3678).

\section*{Data Availability}
The source code, datasets, trained models, and supplementary materials associated with this study are available on our GitHub repository \url{https://github.com/HamidrezaEiv/EquiNO}.

\bibliographystyle{jabbrv_elsarticle_model1-num-names}
\small{
\bibliography{References,literature_troeger,literatur,hartmann-databasis,operatorlearning}
}
\clearpage

\appendix

\section{Ablation Study}

\subsection{Impact of hyperparameters}
\label{app:ablation}
This appendix investigates the sensitivity of EquiNO to key modeling choices, namely the neural network capacity, the number of POD modes used to represent the solution space, and the number of unsupervised training samples used for physics-informed learning. All experiments are conducted on RVE1, and performance is measured using the mean relative $L_2$-norm error of the predicted stress components, 
\begin{equation}
\mu_{\vsigma}
=
\dfrac{1}{d}
\sum_{k=1}^{d}
\dfrac{\left\lVert \sigma_k - \tilde{\sigma}_k \right\rVert_{2}}
     {\left\lVert \sigma_k \right\rVert_{2}} \times 100,
\end{equation}
where $d$ indicates the number of stress components. 
\begin{figure}[ht]
    \centering
    \includegraphics[width=0.9\textwidth]{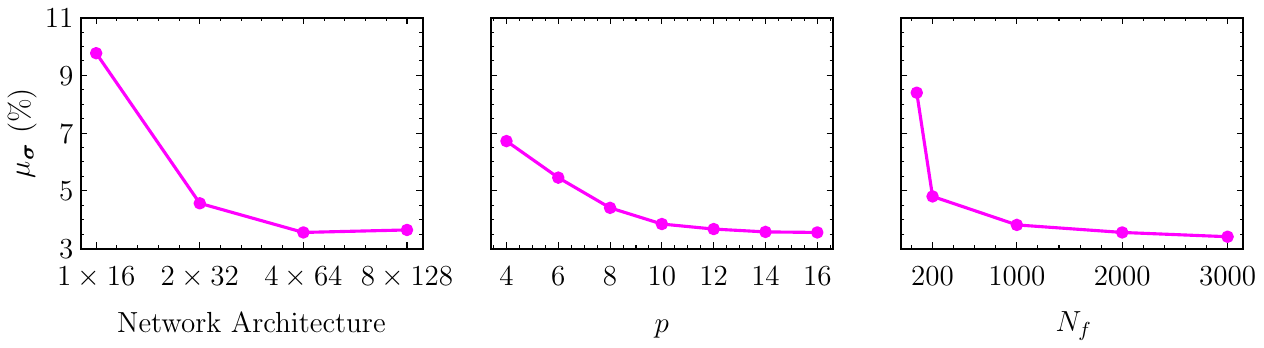}
    \caption{Mean relative $L_2$-norm error of the predicted stress components, $\mu_{\vsigma}$, for different neural network sizes with $p=16$ POD modes (left); for different numbers of POD modes with a fixed network architecture of $4 \times 16$ (middle); and for different numbers of unsupervised training samples (right).}
    \label{fig:ablation}
\end{figure}

\Cref{fig:ablation} (left) shows the effect of increasing the neural network size while keeping the number of POD modes fixed to $p=16$. The network architecture is denoted as $L \times W$, where $L$ is the number of hidden layers and $W$ is the number of neurons per layer and corresponds to the branch networks used for both displacement and stress predictions. The results indicate that increasing the model capacity beyond a moderate size does not further improve accuracy, with errors stabilizing for architectures larger than $4 \times 64$. \Cref{fig:ablation} (middle) illustrates the influence of the number of POD modes $p$ for a fixed network architecture of $4 \times 16$. As expected, increasing $p$ improves accuracy by enriching the reduced bases, but the gains saturate beyond $p \approx 12$. This suggests that a relatively small number of modes is sufficient to accurately capture the solution manifold. Finally, \cref{fig:ablation} (right) shows the effect of the number of unsupervised training samples $N_f$ used for physics-informed learning. Increasing $N_f$ leads to a reduction in the stress error. However, the improvement saturates beyond $N_f \approx 1000$.

\subsection{Role of the divergence-free constraint}
\label{app:dfm}
\Cref{tab:dfm} investigates the impact of enforcing equilibrium through a set of divergence-free POD bases. 
\begin{table}[ht]
    \centering
    \begin{threeparttable}  
    \caption{Relative $L_2$-norm of errors on a test dataset containing 100 samples for each component of displacement vector field and stress tensor field. Results are reported for RVE1.}
    \label{tab:dfm}
    \small
    \begin{tabular}{@{}lcccccc@{}}
    \toprule
    \textbf{Models}     & $u_x(\%)$ & $u_y(\%)$ & $\sigma_{xx}(\%)$ & $\sigma_{yy}(\%)$ & $\sigma_{xy}(\%)$ & $\mu_{\vsigma}(\%)$\\
    \midrule
    EquiNO & 1.1 & 2.01 & 2.35 & 1.51 & 6.82 & \textbf{3.56} \\
    EquiNO (no DFM) & 1.89 & 4.49 & 6.3 & 3.52 & 8.45 & 6.09 \\
    \bottomrule
    \end{tabular}
    \end{threeparttable}  
\end{table}
The standard EquiNO formulation employs divergence-free stress modes (DFM), which ensures that equilibrium is satisfied by construction and allows training with a single stress-based loss. In contrast, EquiNO without divergence-free modes constructs POD modes independently for each stress component, resulting in bases that do not necessarily satisfy equilibrium by construction. Consequently, equilibrium must be enforced softly via the variational loss, and an additional stress loss is required to couple the displacement and stress branches. Loss weighting is performed adaptively using the multi-task learning strategy proposed by \citet{multi-task-learning}.

As shown in \Cref{tab:dfm}, the absence of the divergence-free constraint leads to consistently higher errors, with the mean stress error increasing by more than 70\%. These results highlight that hard enforcement of equilibrium through the choice of basis is not only more accurate but also simplifies optimization by avoiding multi-term loss balancing. This observation is consistent with previous studies showing the benefits of hard-constraining boundary conditions \citep{SUKUMAR2022} and governing physics \citep{CHEN2021} in physics-informed models.

\section{Network architecture and hyperparameters}
\label{app:net}

\paragraph{Fourier Neural Operator (FNO)}
We use an FNO model as a data-driven baseline and a backbone for VINO. The model consists of an input lifting layer followed by two Fourier stages operating on a truncated spectral representation with $8 \times 8$ Fourier modes and a channel width of $16$. To reduce boundary artifacts, zero padding of size $4$ is applied before the spectral layers. Since the microscale fields are evaluated on irregular spatial grids, spectral interpolation is employed to query the learned representation at arbitrary coordinates, enabling the use of an FNO backbone in the finite-element setting. The final projection is performed by a small pointwise MLP that maps the latent features to the output fields.
\begin{table}[ht]
\centering
\caption{Hyperparameters of the FNO used in this work.}
\label{tab:fno_hparams}
\small
\begin{tabular}{@{}ll@{}}
\toprule
\textbf{Hyperparameter} & \textbf{Value} \\
\midrule
Spectral modes in $x$  & $8$ \\
Spectral modes in $y$  & $8$ \\
Channels & $16$ \\
Number of Fourier stages & $2$ \\
Activation & GELU \\
Padding (zero padding) & $4$ \\
Last projection width  & $32$ \\
Output channels & $2$ \\
\bottomrule
\end{tabular}
\end{table}

\paragraph{Operator Transformer (OFormer)}
We use an OFormer baseline consisting of an MLP encoder and a decoder based on cross-attention. The input is first mapped to a latent embedding of dimension $32$ using a depth-$4$ encoder. Predictions at spatial locations are obtained by conditioning a cross-attention decoder on coordinate embeddings, where Fourier embeddings are used to encode the query coordinates. The decoder employs Galerkin-type cross-attention \citep{li2023transformer} and outputs the field values through a small pointwise MLP. This architecture enables evaluation on arbitrary (including irregular) sets of spatial points through coordinate conditioning.
\begin{table}[ht]
\centering
\caption{Hyperparameters of the OFormer baseline used in this work.}
\label{tab:oformer_hparams}
\small
\begin{tabular}{@{}ll@{}}
\toprule
\textbf{Hyperparameter} & \textbf{Value} \\
\midrule
Hidden layers in encoder & $4$ \\
Encoder width & $32$ \\
Decoder latent width & $32$ \\
Attention type & Galerkin cross-attention \\
Activation & GELU \\
Residual connections & enabled \\
Layer normalization & disabled \\
Output channels & $2$ \\
\bottomrule
\end{tabular}
\end{table}

\end{document}